\documentclass[aps,prc,groupedaddress,showpacs,showkeys]{revtex4}
\usepackage{latexsym,epsfig,amsmath,amssymb}
\usepackage{graphicx}
\usepackage{graphics}

\newcommand{\seq}{\begin{subequations}}
\newcommand{\sen}{\end{subequations}}
\newcommand{\eq}{\begin{eqnarray}}
\newcommand{\en}{\end{eqnarray}}
\newcommand{\ra}{\rangle}
\newcommand{\la}{\langle}

\begin{document}
\title{$\eta(1405)$ in a chiral approach based on 
mixing of the pseudoscalar \\ 
glueball with the first radial excitations 
of $\eta$ and $\eta^\prime$} 
\author{Thomas Gutsche, Valery E. Lyubovitskij\footnote{On leave of absence
        from Department of Physics, Tomsk State University,
        634050 Tomsk, Russia}, Malte C. Tichy 
\vspace*{1.2\baselineskip}} 
\affiliation{Institut f\"ur Theoretische Physik, 
Universit\"at T\"ubingen, 
Kepler Center for Astro and Particle Physics, \\ 
Auf der Morgenstelle 14, D-72076 T\"ubingen, Germany
\vspace*{0.3\baselineskip}\\}

\date{\today}

\vskip.5cm 

\begin{abstract}
The supernumerous $\eta(1405)$ is considered a strong candidate for 
the pseudoscalar glueball. In a phenomenological chiral approach, 
we consider a mixing scenario of bare pseudoscalar $n \bar n$ and 
$s \bar s$ quarkonia with the  glueball. We study the decay 
properties and point out the peculiarities of this scenario in order 
to support the possible identification of the $J^{\rm PC} = 0^{-+}$ 
glueball.
\end{abstract}
\maketitle 
\newpage
\section{Introduction}
The problem of glueballs is paradigmatic for hadronic physics and has 
been under intense theoretical and experimental analysis for many 
years (see e.g. Ref.~\cite{Donoghue:1980hw}-\cite{Mathieu:2008up}). 
It is still an important ongoing topic since the existence and the 
properties of glueballs are related to very fundamental properties 
of QCD, such as chiral symmetry breaking and the non-abelian nature 
of the gauge group SU(3). The lowest-lying glueball has been predicted 
by theoretical calculations to have scalar quantum 
numbers~\cite{Chen:2005mg,Meyer:2004jc}.

The case for higher glueball excitations, such as the pseudoscalar one, 
is more difficult: So far, no unquestionable candidate has been observed. 
The $\eta(2225)$, with appropriate quantum numbers, is excluded by the 
measured decay pattern and weak coupling to 
gluons~\cite{Bai:1990hk,Heusch:1991sw}. Regarding the $X(1835)$, 
there is a current debate about its nature, and it has been as well 
discussed as a glueball candidate~\cite{Kochelev:2005tu}. 
However, the conclusions are not definitive, and further investigation 
is needed here. The possibly non-vanishing gluonium content of the 
ground state $\eta$ and $\eta^\prime$ mesons is discussed
 in~\cite{Ambrosino:2006gk,Escribano:2007cd,Thomas:2007uy,Cheng:2008ss}.
In the mass region of the first radial excitation of the $\eta$ and 
$\eta^\prime$ mesons, a supernumerous candidate, the $\eta(1405)$ 
has been observed. An excellent review on the experimental status
of the $\eta(1405)$ is given in Ref.~\cite{Masoni:2006rz}.
This state lies considerably 
lower than the lattice QCD predictions, which suggest a glueball 
around 2.5~GeV \cite{Gabadadze:1997zc,Morningstar:1999rf}. 
On the other hand, there are compelling arguments for the pseudoscalar 
glueball being approximately degenerate in mass with the scalar 
glueball~\cite{Faddeev:2003aw}.
Even the scenario that a pseudoscalar glueball is lower in mass 
than the scalar one is recently discussed in Ref.~\cite{He:2009sb}. 
Therefore, more phenomenological 
estimates are needed to resolve this issue, and mixing has to be 
included as an important ingredient in a model since the three observed 
isoscalar states, $\eta(1295)$, $\eta(1405)$ and $\eta(1475)$ are close 
in mass.  
Besides the 'standard scenario', where the first radial excitations of the $\eta$-states
are supposed to reside in the 1300 - 1500 MeV mass region, other structure interpretations
are also discussed. For example, as originally suggested in~\cite{Fariborz1} and
recently discussed in\cite{Fariborz2}, an interpretation of the heavy $\eta$ states as
four-quark-states including mixing with the conventional quarkonia states is also feasible.

One of the key features to disentangle the properties of pseudoscalar mesons possibly
mixed with a glueball is a good understanding of their strong decay patterns.
The most convenient language for the treatment of light 
hadrons at small energies was elaborated in the context of 
Chiral Perturbation Theory 
(ChPT)~\cite{Weinberg:1978kz,Gasser:1983yg,Ecker:1988te},  
the effective low--energy theory of the strong interaction.  An extension 
of this approach above the chiral scale of about 1 GeV has been shown in 
the past to also reproduce the main dynamical properties of meson 
resonances.  Hence it constitutes a useful phenomenological tool for 
the estimate of decay widths and ratios. This work is a continuation of 
a series of papers~\cite{excited_mesons,Giacosa:2005qr,Giacosa:2005bw,%
Giacosa:2006tf}, where we 
analyzed the decays of scalar (including mixing with the scalar glueball), 
tensor (including mixing with the tensor glueball), pseudoscalar and 
vector mesons using a phenomenological chiral approach.

In the present paper, we consider the possible mixing of the ground state 
pseudoscalar glueball with the first radial excitations of the $\eta$ and 
$\eta^\prime$-meson. The main contributions to the physical states are 
assumed to be $\eta(1295) \approx n \bar n, \eta(1405) \approx G, 
\eta(1475) \approx s\bar s$. We give constraints on the nature of the 
glueball and on the resulting decay patterns. Furthermore, the available,
sometimes contradictory,
experimental data are confronted with the predictions of the model. 
A glueball interpretation of the $\eta(1405)$ is consistent with the 
present, sparse experimental data, but the assignment is still far from 
unique. Note that an analysis~\cite{Li:2007ky} of data on
$J/\psi (\psi^\prime )$ decays into
vector and pseudoscalar mesons also concludes that the $\eta(1405)$ has a dominant
pseudoscalar glueball component.
Forthcoming data from the planned experiments at BES-III,
COMPASS (see for example the overview of~\cite{Crede:2008vw}) and at the 
upgrade facility FAIR at GSI~\cite{Bettoni:2005ut} 
might allow a full quantitative test of the mixing scenario.

The paper is structured as follows. In Sec.~\ref{sec:theo} 
we discuss our formalism.  The results of the calculations are given in 
Sec.~\ref{sec:results}. Conclusions 
and outlook on future experimental and theoretical directions are  
presented in~\ref{last}. 

\section{Theoretical foundations} \label{sec:theo}
\subsection{The Lagrangian}
We employ a chiral Lagrangian to describe the coupling of the pseudoscalar 
excitations to the decay channels: pseudoscalar and vector meson (PV), 
three pseudoscalar mesons (PPP), pseudoscalar and scalar meson (PS). 
The Lagrangian including the chiral fields ($U$), the scalar (${\cal S}$), 
the vector (${\cal V}$) and the excited pseudoscalar (${\cal P}^\ast$) 
mesons, defined in Appendix A, reads as 
\eq
{\cal L} &=&   \frac{F^{2}}{4} \left\la D_{\mu }U\,D^{\mu} 
U^{\dagger }+\chi _{+} \right\ra + {\cal L}_{\mathrm{mix}}^{\cal P} 
+  \frac{1}{2}  \left\la \nabla_\mu {\cal P^{\ast}} 
\nabla^\mu {\cal P^{\ast}} - M_{{\cal P^{\ast}}}^2 {\cal P^{\ast}}^2 \right\ra 
+ {\cal L}_{\mathrm{mix}}^{\cal P^{\ast}}   \nonumber \\ 
&+& \frac{1}{2} \left\la \nabla_\mu {\cal V}^{\mu\nu} 
\nabla^\rho {\cal V}_{\rho\nu} 
- \frac{1}{2} M_{\cal V}^2 {\cal V}_{\mu\nu} {\cal V}^{\mu\nu} 
\right\ra + {\cal L}_{\mathrm{mix}}^{\cal V}   
+ \frac{1}{2} \left \la \nabla_\mu {\cal S} \nabla^\mu {\cal S} 
- M_S^2 {\cal S}^2\right\ra \nonumber \\
&+& c_{_{P^\ast PV}} \left\langle {\cal V}_{\mu\nu}
\left[ u^\mu, \nabla^\nu {\cal P}^* \right] \right\rangle  
+ i c_{_{P^\ast PPP}} \left\langle {\cal P}^* \chi_- \right\rangle 
+ c_{_{P^{\ast} PS,1}} \left \langle {\cal S} \left\{ 
\nabla_{\mu}{\cal P}^*, u^{\mu} \right\} \right\rangle 
+ c_{_{P^{\ast} PS,2}} 
\left \langle {\cal S} \left\{ {\cal P}^\ast , \chi_- \right\} 
\right\rangle \; .  \label{L_eff}
\en
Here the symbols $\la \, \cdots \, \ra$, 
$[ \, \cdots \, ]$ and $\{ \, \cdots \, \}$ occurring in Eq.~(\ref{L_eff}) 
denote the trace over flavor matrices, commutator, and anticommutator, 
respectively. The constants $c_{_{P^\ast PV}}, c_{_{P^\ast PPP}}, 
c_{_{P^{\ast} PS,1}},  c_{_{P^{\ast} PS,2}}$ define the couplings of the 
excited pseudoscalar fields to the decay channels
PV, PPP and PS, respectively. The terms $\cal L_{\mbox{mix}}^P$, 
$\cal L_{\mbox{mix}}^{P^\ast}$ and $\cal L_{\mbox{mix}}^V$ describe 
the mixing between the octet and singlet of the pseudoscalar, 
excited pseudoscalar and vector mesons, respectively. Due to the axial 
anomaly, we also encode an additional contribution to the mass of the 
$\eta^0$~\cite{Giacosa:2005qr}.

We use the standard notation for the basic blocks of the ChPT 
Lagrangian~\cite{Weinberg:1978kz}: $U=u^{2}=\exp (i{\cal P}\sqrt{2}/F)$ 
is the chiral field collecting
pseudoscalar fields in the exponential parametrization, 
$D_{\mu }$ and $\nabla_{\mu }$ denote the chiral and gauge-invariant 
derivatives, $u_{\mu }=iu^{\dagger }D_{\mu
}Uu^{\dagger }$, \hspace*{0.2cm} $\chi _{\pm }=u^{\dagger }\chi u^{\dagger
}\pm u\chi ^{\dagger }u,\hspace*{0.2cm}\chi =2B(s+ip),\,\,\,s={\cal M}%
+\ldots \,$ and $F_{\mu \nu }^{+}\,=\,u^{\dagger }F_{\mu \nu }Qu+uF_{\mu
\nu }Qu^{\dagger }\,,$ where $F_{\mu \nu }$ is the stress tensor of the
electromagnetic field; $Q={\rm diag}\{2/3,-1/3,-1/3\}$ and ${\cal M}={\rm %
diag}\{\hat{m},\hat{m},m_{s}\}$ are the charge and the mass matrix of
current quarks, respectively (we restrict to the isospin symmetry limit with 
$m_{u}=m_{d}=\hat{m}$); $B$ is the quark vacuum condensate parameter. Then
the masses of the pseudoscalar mesons in the leading order of the chiral
expansion are given by $M_{\pi }^{2}=2\hat{m}B\,\,,M_{K}^{2}=(\hat{m}%
+m_{s})B\,\,,\,M_{\eta ^{8}}^{2}=(2/3)(\hat{m}+2m_{s})B\,.$
The glueball configuration is not yet included in this Lagrangian, it will be 
introduced further on as a flavor singlet which mixes with the 
corresponding excited $\eta^{\ast}$-states. 

We intend to employ the Lagrangian to the tree-level
calculation of the strong decays of radially excited pseudoscalar mesons. 
At the energy scale of interest, $E \sim M_{{\cal S}}~\sim~1.5$ GeV, 
a calculation of loops and an application of the power counting rules
are not rigorously justified. The aim of the present approach is therefore 
a phenomenological study of pseudoscalar meson physics, for which 
a tree-level calculation represents a useful analysis.

\subsection{The mixing scenario}

We treat $\eta(1295)$ and $\eta(1475)$ as two members of the nonet of 
radial pseudoscalar excitations. The glueball $G$ is added as 
a flavor singlet: 
\eq 
{\cal L}_G=\frac 1 2  \partial_\mu G \partial^\mu G 
- \frac{1}{4} M_{G}^2 G^2  + {\cal L}_{G,\mbox{decay}} 
\en  
where 
\eq 
{\cal L}_{G,\mbox{decay}} =  c_{_{G PV}} \left\langle {\cal V}_{\mu\nu}
\left[ u^\mu, \partial^\nu G \right] \right\rangle  
+ i c_{_{G PPP}} \left\langle G \chi_- \right\rangle 
\nonumber 
+ c_{_{G PS,1}} \left \langle {\cal S} \left\{ \partial_{\mu} 
G, u^{\mu} \right\} \right \rangle + c_{_{G PS,2}} 
\left \langle {\cal S} \left\{ G , \chi_- \right\} 
\right\rangle  
\en 
contains the coupling of the pseudoscalar glueball singlet to its decay 
channels in analogy to the excited meson octet. 

To incorporate mixing, we replace the diagonal term with the bare masses
of $\eta_{nn}$, $\eta_{ss}$ and of the unmixed glueball in the Lagrangian 
with the following non-diagonal mixing term given
in general form:
\eq 
\tilde M^2 = \left(\begin{array}{ccc}
M_{n \bar n}^2 & \sqrt 2 f r &  \epsilon  \\
 \sqrt 2 f r  &  M_{g g}^2 &   f  \\
\epsilon & f & M_{s \bar s}^2
 \end{array} \right)\,,
\en 
where $\epsilon$ denotes the coupling between the $s \bar s$ and $n \bar n
= (u \bar u + d \bar d) /\sqrt{2}$ flavor configurations. 
This parameter is set to zero in the following discussion. One reason for doing
so is traced to an analogous treatment in the scalar
sector~\cite{Amsler:1995td,Lee:1999kv}, where it was argued
that the mixing between excited quarkonia states is suppressed relative to the quarkonia-glueball
mixing mechanism. Also, the mass degeneracy between the $\eta (1295)$ and the $\pi (1300)$
suggests a nearly ideal mixing situation for the quarkonia configurations,
implying in turn a strongly suppressed mixing between the quarkonia states.  
The parameter $f$ denotes the mixing of the glueball with the quarkonia 
states. A possible deviation from the case of flavor symmetric mixing is 
expressed by the parameter $r$ with $r \ne 1$. 
In our case, however, we will work in the 
limit $r=1$; the same limit is also approximately
fulfilled in the scalar case \cite{Giacosa:2005qr}.

After diagonalization of $\tilde M$ the physical mass matrix reads as 
\eq
M^2 = U \tilde M^2 U^T = 
\left(\begin{array}{ccc}
M_{\eta_1}^2 & 0  & 0  \\
 0 & M_{\eta_2}^2 &   0  \\
0  & 0 &  M_{\eta_3}^2
\end{array} \right) \label{massesreal} \,, 
\en 
where $U$ is the mixing matrix relating the physical 
states $(\eta_1, \eta_2, \eta_3)$ to the bare states 
$(\eta_{n \bar n}, \eta_{g g}, \eta_{s\bar s})$ as: 
\eq 
U \left( 
\begin{array}{c} 
\eta_{n \bar n} \\ 
\eta_{g g}  \\ 
\eta_{s \bar s} 
\end{array} 
\right)  = 
\left( 
\begin{array}{c} 
\eta_1 \\ 
\eta_2 \\ 
\eta_3 
\end{array} 
\right) \; .  
\en  
The mixing strength $f$ is one parameter of interest which will be 
varied in the following discussion. Furthermore, we will vary the 
relative decay strength of the bare glueball with respect to the 
decay strength of the quarkonia states. Thus we set up the relation
\eq 
c_{G}=g c_{{\cal P}^\ast} 
\en 
applicable to
every coupling constant. The limit $g=0$ indicates no direct decay of
the glueball component, 
interference effects will be observed when changing the sign, 
e.g. from $g=-1$ to $g=+1$.

\section{Results}\label{sec:results} 

In the case of decays to the PV channel we used the results obtained 
within the $^3P_0$-model by Barnes et al.~\cite{Barnes:1996ff,Barnes:2002mu} 
to fit the coupling strength to $c_{_{P^\ast PV}} = 4.95$
GeV$^{-1}$~\cite{excited_mesons}.
This procedure allows the computation of decay widths 
in physical units. For the three-pseudoscalar channel and the 
scalar-pseudoscalar channel no absolute value can be given and 
we restrict ourselves to the ratios of rates. 

\begin{figure}
\begin{center} 
\includegraphics[width=9.5cm,angle=0]{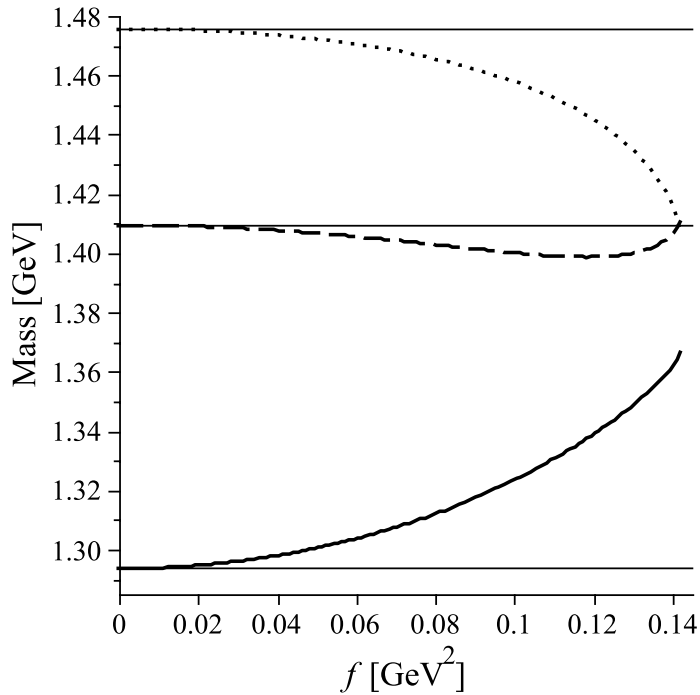}
\end{center}
\caption{Bare masses $M_{n\bar n}$ (solid line), 
$M_{s \bar s}$ (dotted line), 
$M_{gg}$ (dashed line). The bare masses shown here 
generate for different values of $f$ the same physical masses 
$M_{\eta_1}, M_{\eta_2}, M_{\eta_3}$. 
The bare mass of a pseudoscalar glueball is constrained to 
$\approx 1.39-1.41$~GeV.  } \label{baremassesforphys}
\end{figure}

In order to obtain the given physical masses, 
one is restricted to mixing strengths smaller than about 
$f\approx 0.15$~GeV$^2$. A moderate value for the mixing strength however
does not induce considerable mass shifts of the bare values,
as can be seen in 
Fig.~\ref{baremassesforphys}. Therefore, a three-state mixing scenario cannot 
explain the discrepancy in the mass values, which persists between the 
predictions of Lattice QCD and the examined candidate. 

A glueball-free scenario gives consistent results in decay when $\eta(1295)$ 
is interpreted as a dominant $n\bar n$ quarkonium configuration and 
$\eta(1475)$ is seen as the $s \bar s$ state~\cite{excited_mesons}. 
This observation also motivates a mixing scenario with a small mixing strength.

\subsection{Decay to $K \overline{K^\ast}$}

\begin{figure}
\begin{center} 
\includegraphics[width=9.5cm,angle=0]{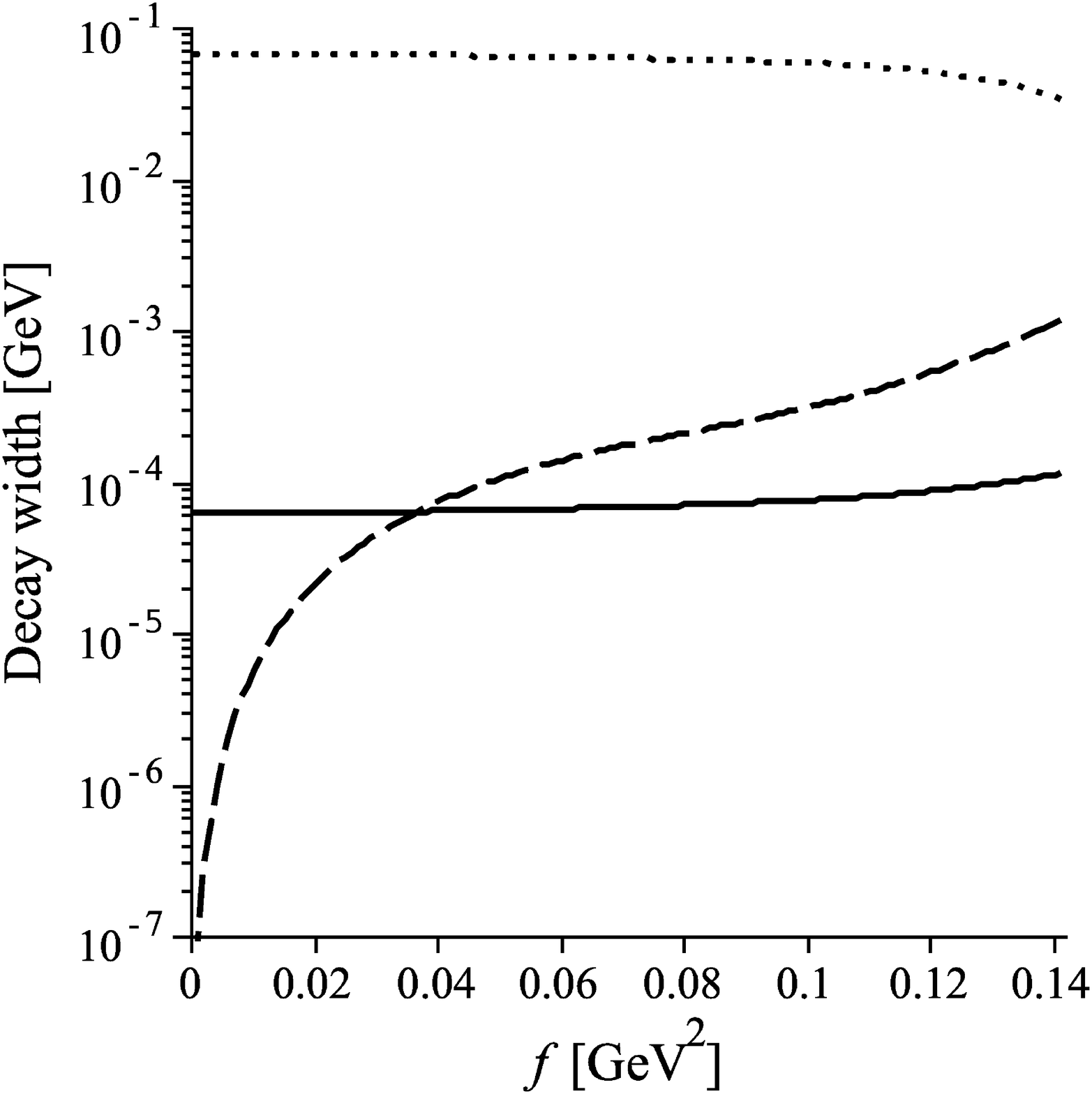} 
\end{center}
\caption{Decay widths to $K \overline K^\ast$ of $\eta(1295)$ 
(solid line), $\eta(1475)$ (dotted line), $\eta(1405)$ (dashed line). 
The decay widths of $\eta(1475)$ and $\eta(1295)$ do not change 
considerably in the range of $f$, however the decay width of $\eta(1405)$ 
changes over several orders of magnitude. }\label{threekkstar}
\end{figure}

In the chiral approach the pure glueball configuration does not decay to 
$K \overline{K^\ast}$, but only via mixing. The resulting decay width 
therefore depends very strongly on the mixing strength. The decay 
widths of the respective $\eta$-states as a function of the mixing strength
are shown in Fig.~\ref{threekkstar}. 
The decay width $\Gamma (\eta (1475) \to K \overline K^\ast)$ lies
in the range of values from 34 MeV to 68 MeV. 
Concerning data on $\eta (1405)$ the experimental situation
is contradictory. Although many experiments do not observe
$\eta (1405) \to K \overline K^\ast$~\cite{Masoni:2006rz},
the analysis of E852~\cite{Adams:2001sk} indicates a ratio of
\eq 
\mbox{Br}(\eta (1405) \to K^\ast \overline K)/\mbox{Br}(\eta (1475) 
\to K^\ast \overline K)=0.16\pm 0.04 
\en 
with statistical error only, 
but where the systematic error is expected to be large.
Last ratio cannot be explained in the present framework, even when
additional quarkonia mixing is included. 
An experimental clarification of the possible $K\overline K^\ast $
decay mode of the $\eta (1405)$ is obviously very useful.

\subsection{Decay to three pseudoscalars}

We consider the two kinematically allowed three-body decays to $\pi \pi \eta$ 
and to $K \overline K \pi$.
Note that these channels refer to the direct three-body
decay modes and not to the final state fed for example by intermediate 
$f_0(980) \eta$ or $K^\ast \overline K$ decay channels. 
The glueball as a flavor-singlet state can decay into $\eta \pi \pi$ as well 
as into $\pi K \overline K$, therefore an additional dependence on $g$, 
indicating a contribution of direct glueball decay, arises. 
In Fig.~3 we show the decay widths of all three 
resonances to $\pi \pi \eta$ for different discrete values of the 
direct glueball decay strength, $g=-1,0,1$. The decay widths are given 
in arbitrary units, but the relative rates are a prediction of the model. 
It is easy to see that the decay width strongly depends both on the direct 
glueball decay and on the mixing strength.

The decay of $\eta(1475)$ is strongly suppressed for 
small values of $f$, even when the direct glueball decay is very 
large. In the case of a vanishing direct glueball decay, it can only 
decay via its $n \bar n$-component, which is small. Interference effects 
are important and lead to very different behavior when considering $g=-1$ 
instead of $g=1$. In the case of the $\eta(1295)$, for a very large mixing 
strength, destructive interference leads to a vanishing decay width.

The same analysis can be repeated for the decay into $K \overline K \pi$.
The results are shown in Fig.~4.
The decay pattern is comparable to the previous case. 
Since the component $\eta_{s\bar s}$ can now also feed the decay channel 
$K \overline K \pi$, the decay width of $\eta (1475)$  
does not vanish in case of $f = 0$ and $g = 0$.

\begin{figure}
\begin{center} 
\includegraphics[width=7.5cm,angle=0]{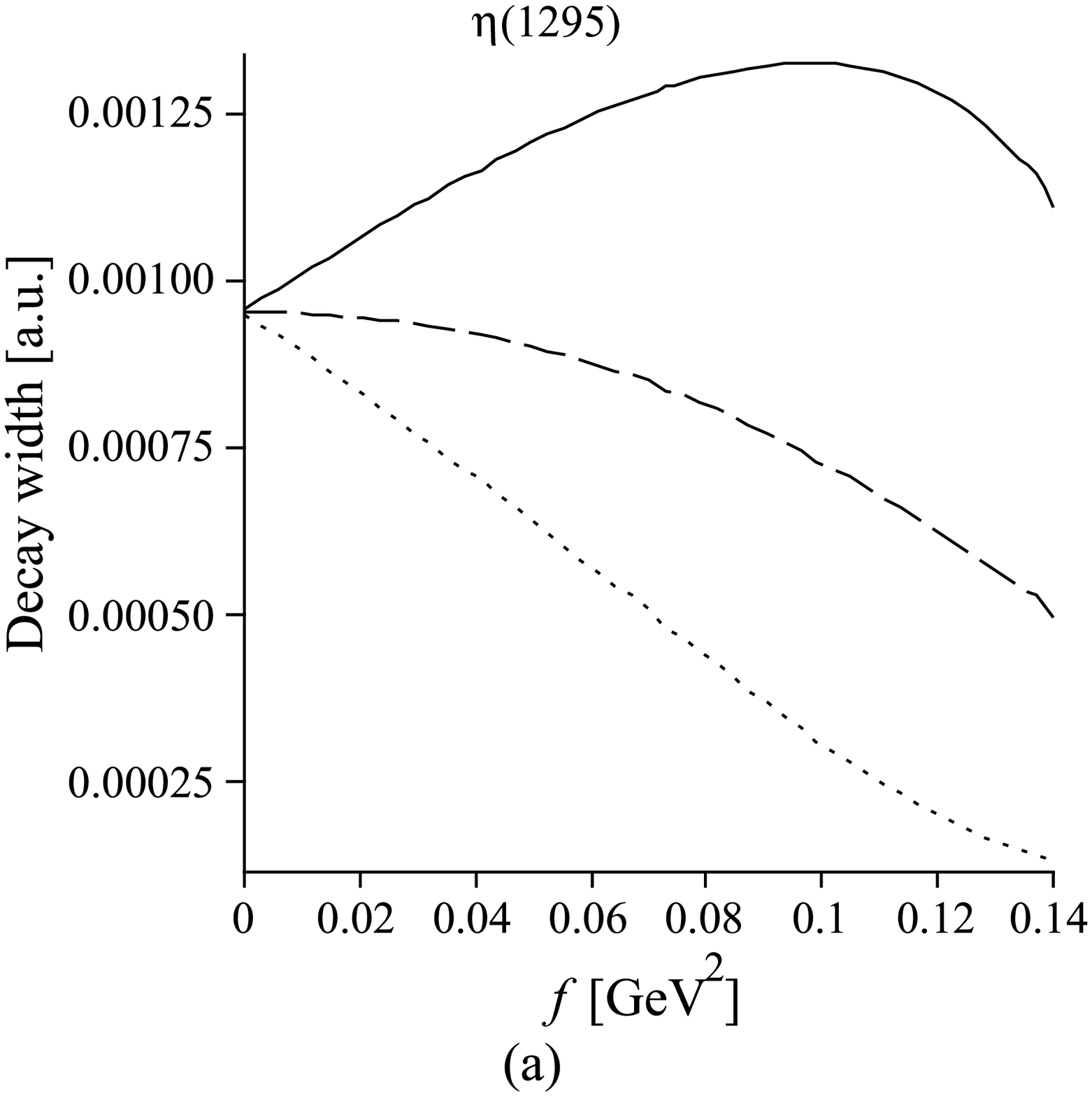} 
\includegraphics[width=7.5cm,angle=0]{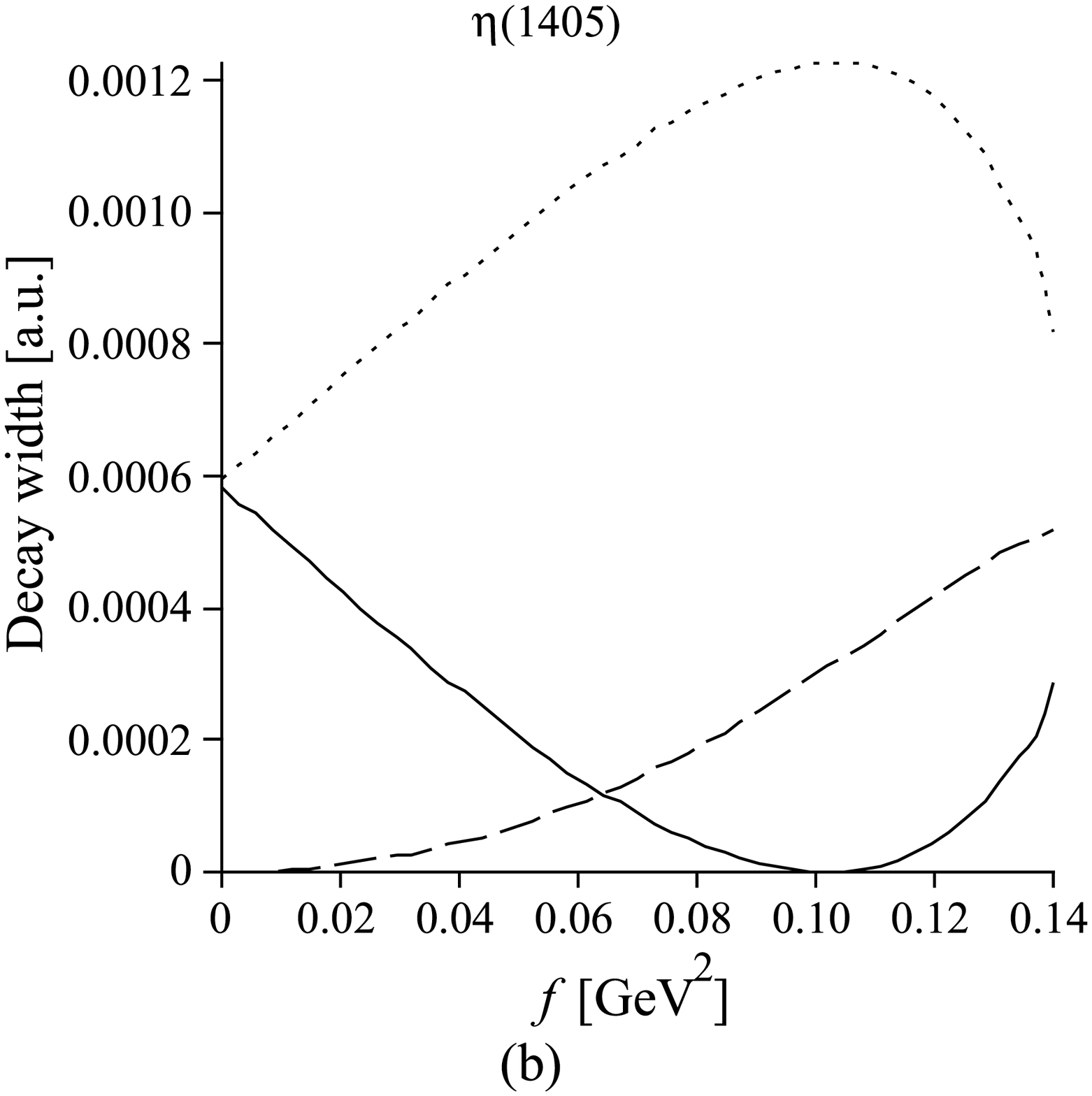} \\
\includegraphics[width=7.5cm,angle=0]{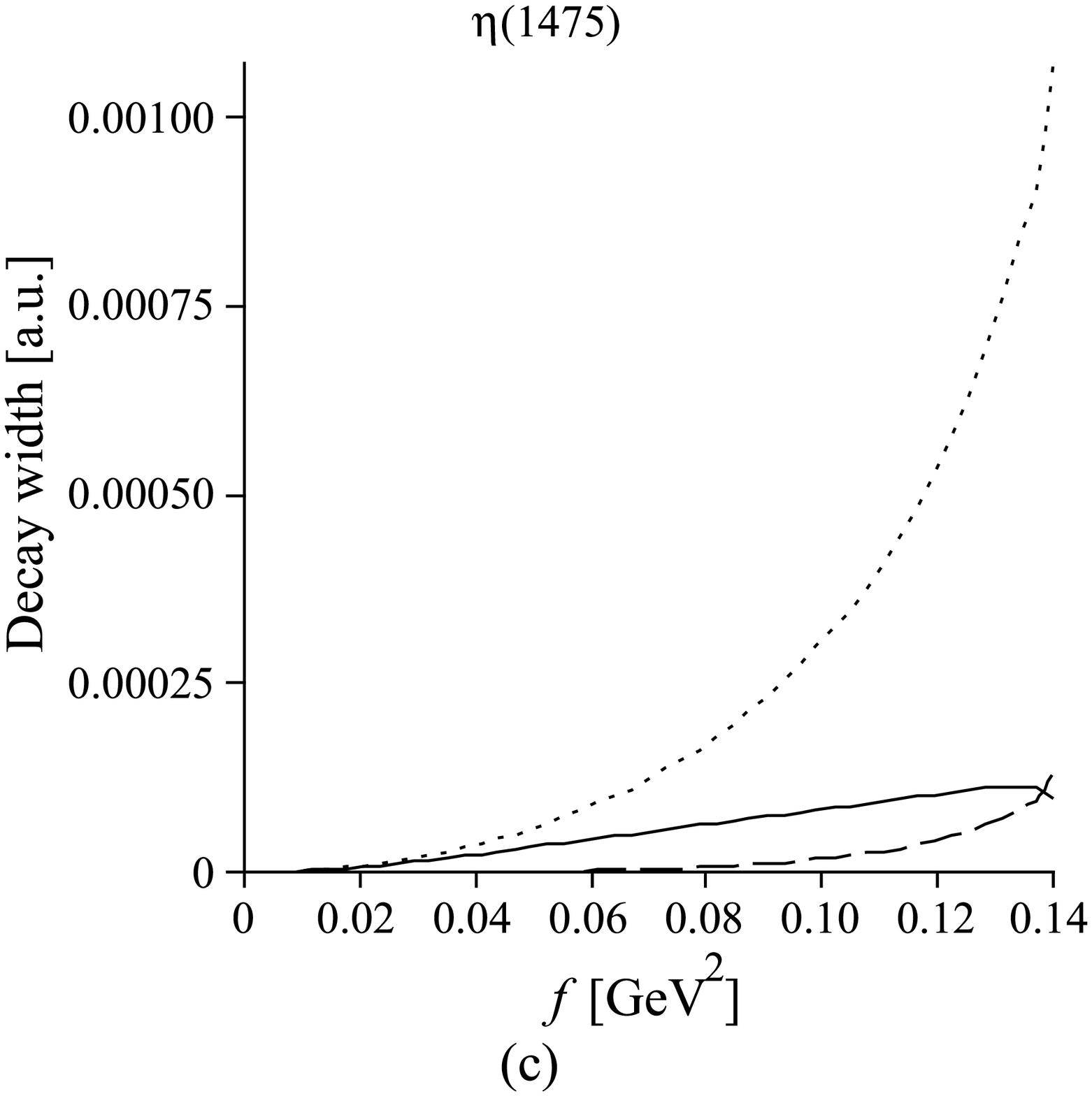} 
\caption{Decay widths to $\pi \pi \eta$ with arbitrary, absolute 
normalization. The decay width is shown for three different 
relative glueball decay strengths: $g=0$ (no direct glueball decay, 
dashed line), 
$g=1$ (dotted line), $g=-1$ (solid line).}   
\end{center}
\label{fig:threethree}
\end{figure}
\begin{figure}
\begin{center} 
\includegraphics[width=7.5cm,angle=0]{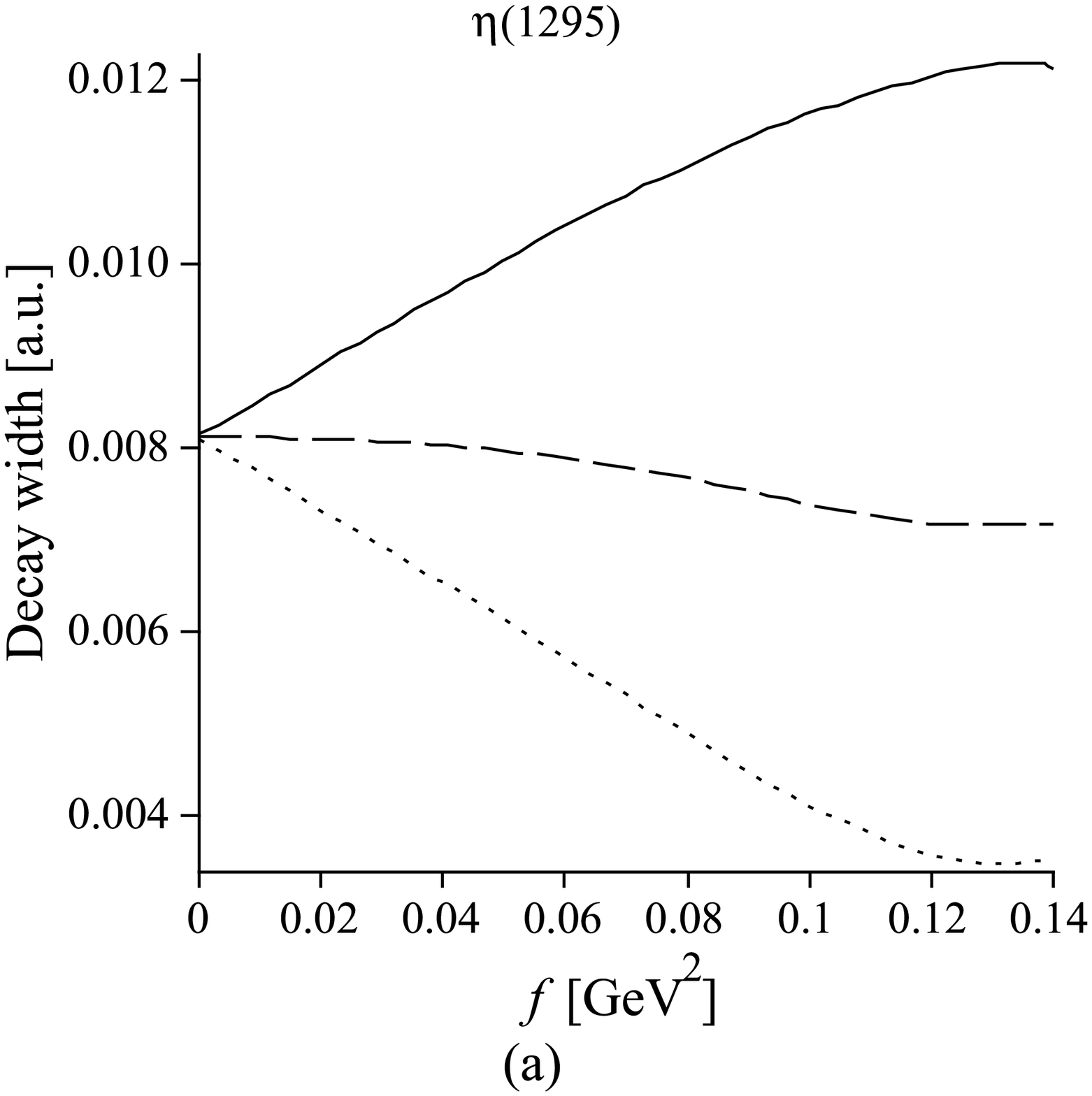} 
\includegraphics[width=7.5cm,angle=0]{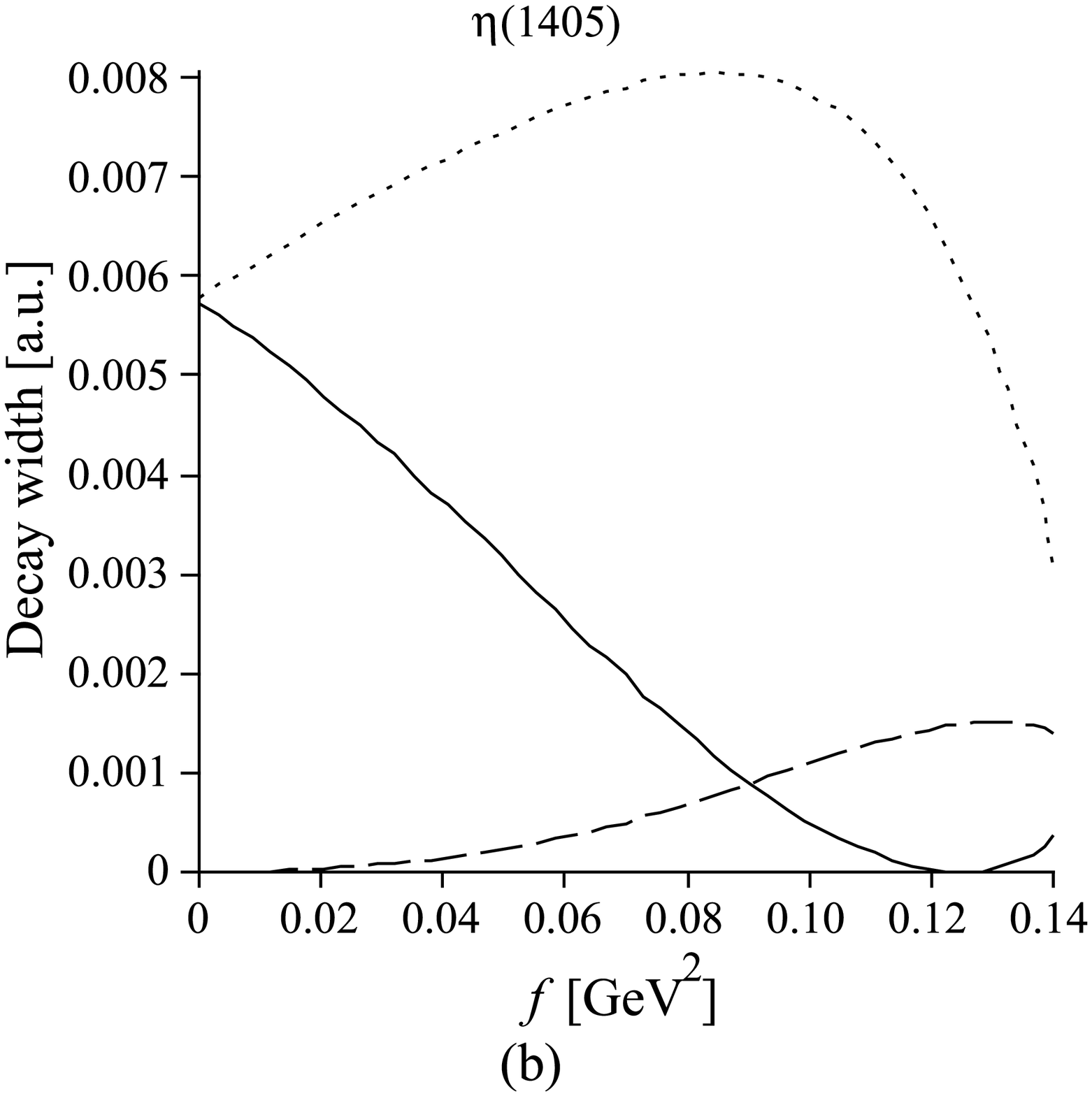} \\
\includegraphics[width=7.5cm,angle=0]{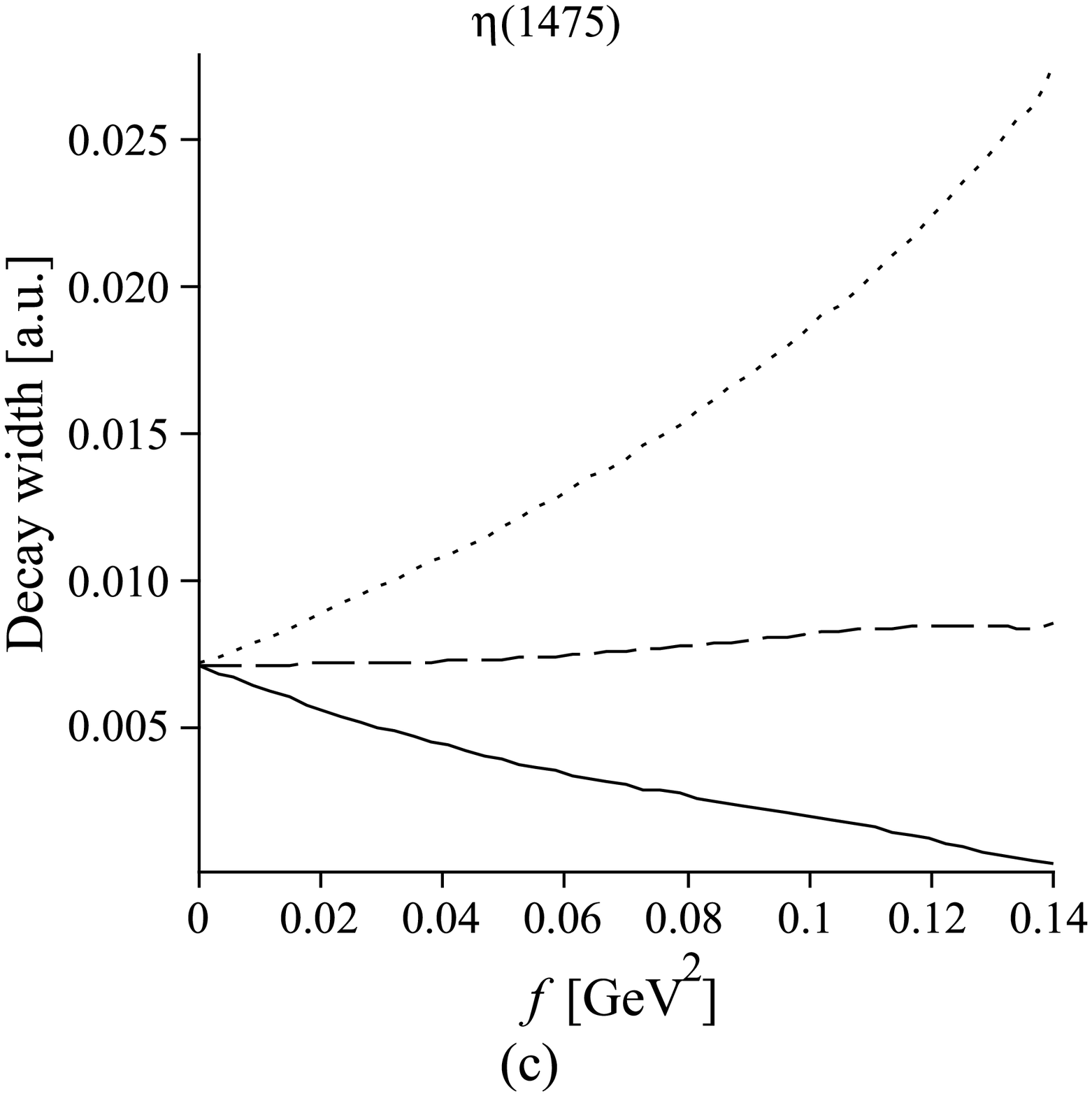} 
\caption{Decay widths to $\pi K \overline K$.
The normalization and legend is identical to the 
one of Fig.~3. A strong admixture of $n \bar n$ 
to the $s \bar s$ state leads to a smaller decay width of the 
$\eta(1475)$ for $g=0$. } 
\end{center}
\label{threethree2}
\end{figure}
\begin{figure}
\begin{center} 
\includegraphics[width=7.5cm,angle=0]{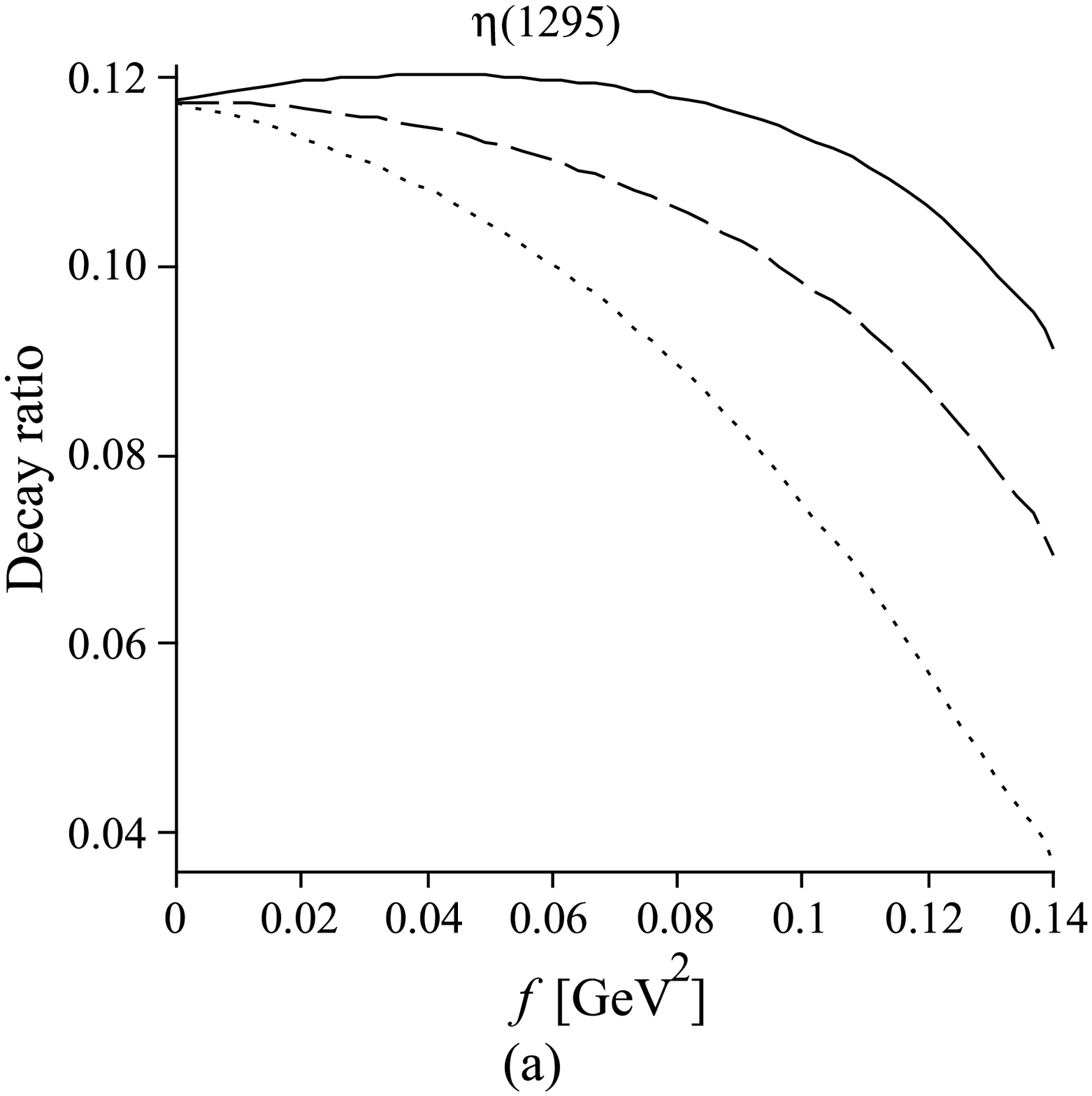} 
\includegraphics[width=7.5cm,angle=0]{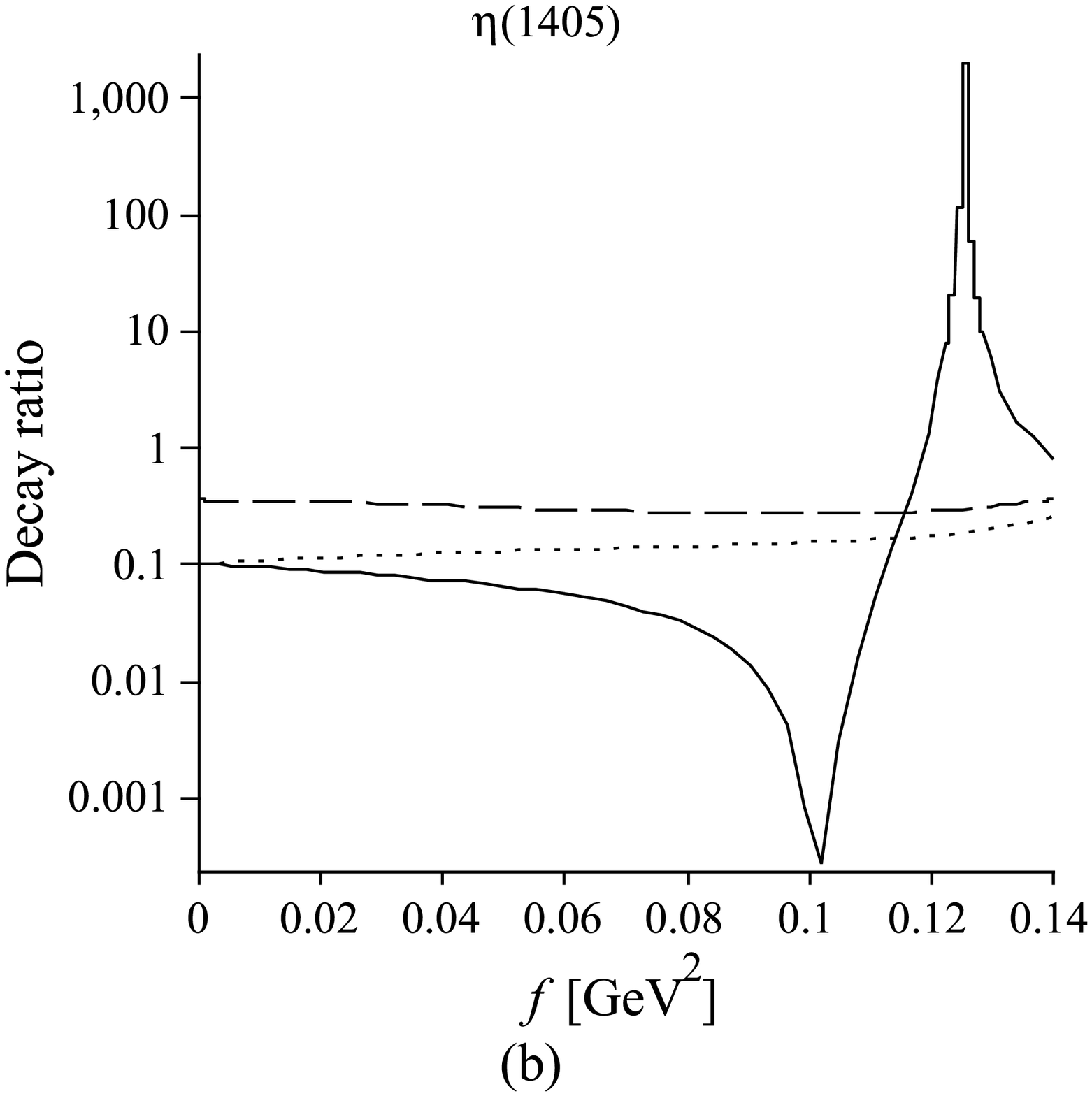} \\
\includegraphics[width=7.5cm,angle=0]{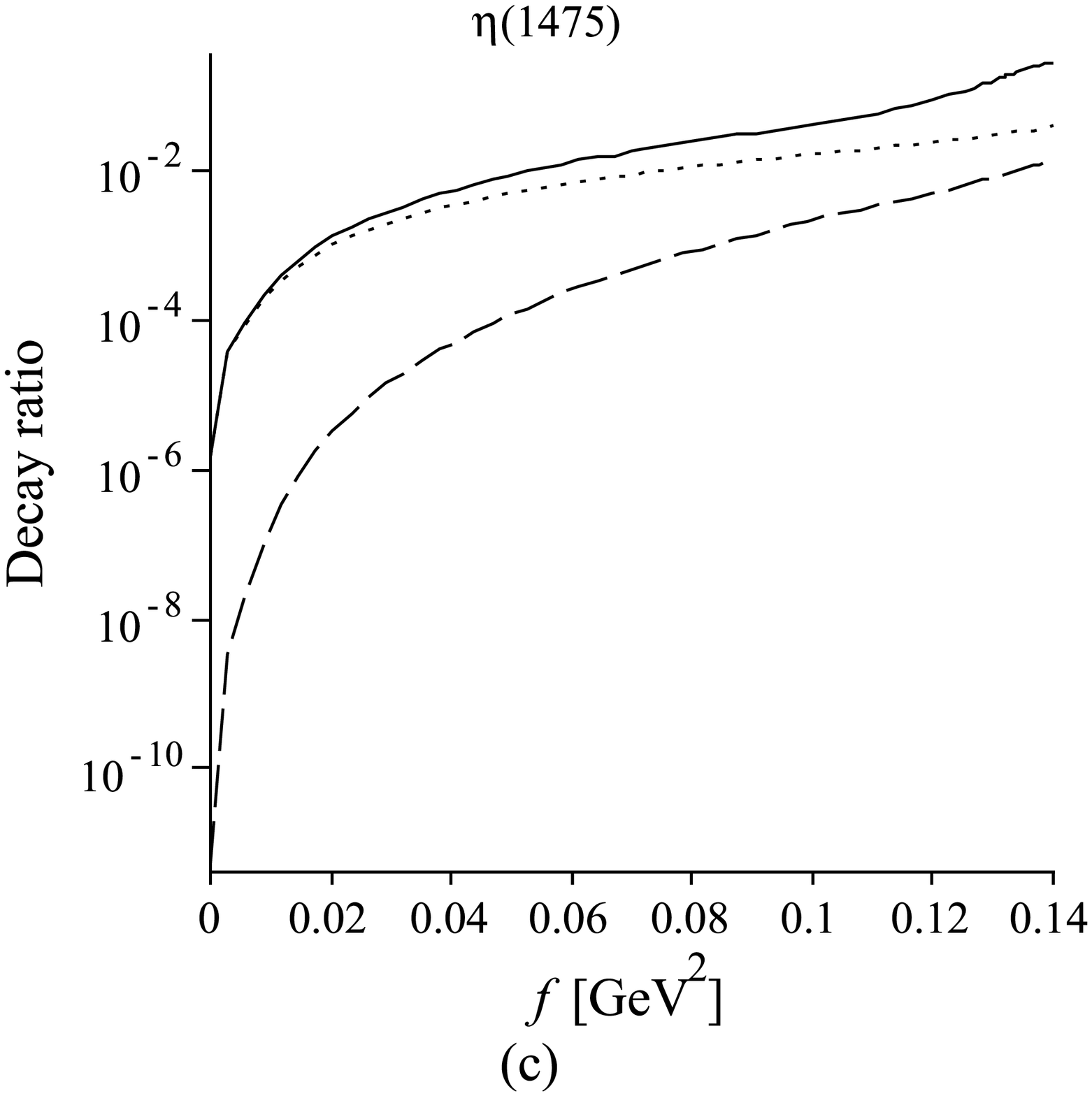} 
\caption{Decay ratios 
$\frac{\Gamma(\eta_i \rightarrow \pi K \overline K)}{\Gamma(\eta_i \rightarrow 
\pi \pi \eta)}$. We consider the cases $g=0, g=1, g=-1$ 
(legend analogous to previous two figures). 
Due to the wide range of values the plots for $\eta(1405)$ and $\eta(1475)$ 
are shown in logarithmic scale. Compared to the cases of $\eta(1475)$ 
and $\eta(1405)$, the $f$ dependence of the decay ratio for $\eta(1295)$ 
is less pronounced.} \end{center}
\label{fig:threethreeratio}
\end{figure}

Since the decay strength - although unknown - is the same for both three-body
decays analyzed here, we can consider the decay ratio 
\eq 
\frac{\Gamma(\eta_i \rightarrow \pi K \overline K)}{\Gamma(\eta_i 
\rightarrow \pi \pi \eta)}. 
\en 
Its dependence on the various choices 
of parameters is indicated in Fig.~5. Similarly, the predictions for the
ratios
$\Gamma(\eta_i \rightarrow \pi K \overline K)/ 
\Gamma(\eta_j \rightarrow \pi K \overline K)$
and 
$\Gamma(\eta_i \rightarrow \pi \pi \eta)/ 
\Gamma(\eta_j \rightarrow \pi \pi \eta)$
are independent of the coupling strength.
We note that the decay ratios of Fig.~5 for $\eta(1295)$ and $\eta(1405)$  
remain rather unaffected when taking a reasonable glueball-admixture, 
independent of the direct decay strength. For $\eta(1475)$, 
the situation is different: Since $\eta \pi \pi $ is totally suppressed 
for a pure $s \bar s$ configuration, even small $n \bar n$-admixtures 
lead to a non-vanishing decay width and a strong change in the ratio. 

\subsection{Decay to scalar and pseudoscalar mesons} 

Several decay channels to scalar and pseudoscalar mesons open in the 
final state: The most important ones are $\kappa (K^\ast_0 (800)) K$, 
$\sigma (f_0 (600)) \eta$, $f_0(980) \eta$, $a_0 (980) \pi$.
We assume that flavor symmetry remains unbroken in this 
scenario. It can however be shown that a flavor-symmetry-breaking term 
does not induce dramatic changes in the decay pattern. Since we are 
rather interested in the dependence on the mixing strength and the 
direct glueball decay strength, we omit an analysis of this effect here. 

\subsubsection{Omission of direct glueball decay}

\begin{figure}
\begin{center} 
\includegraphics[width=7.5cm,angle=0]{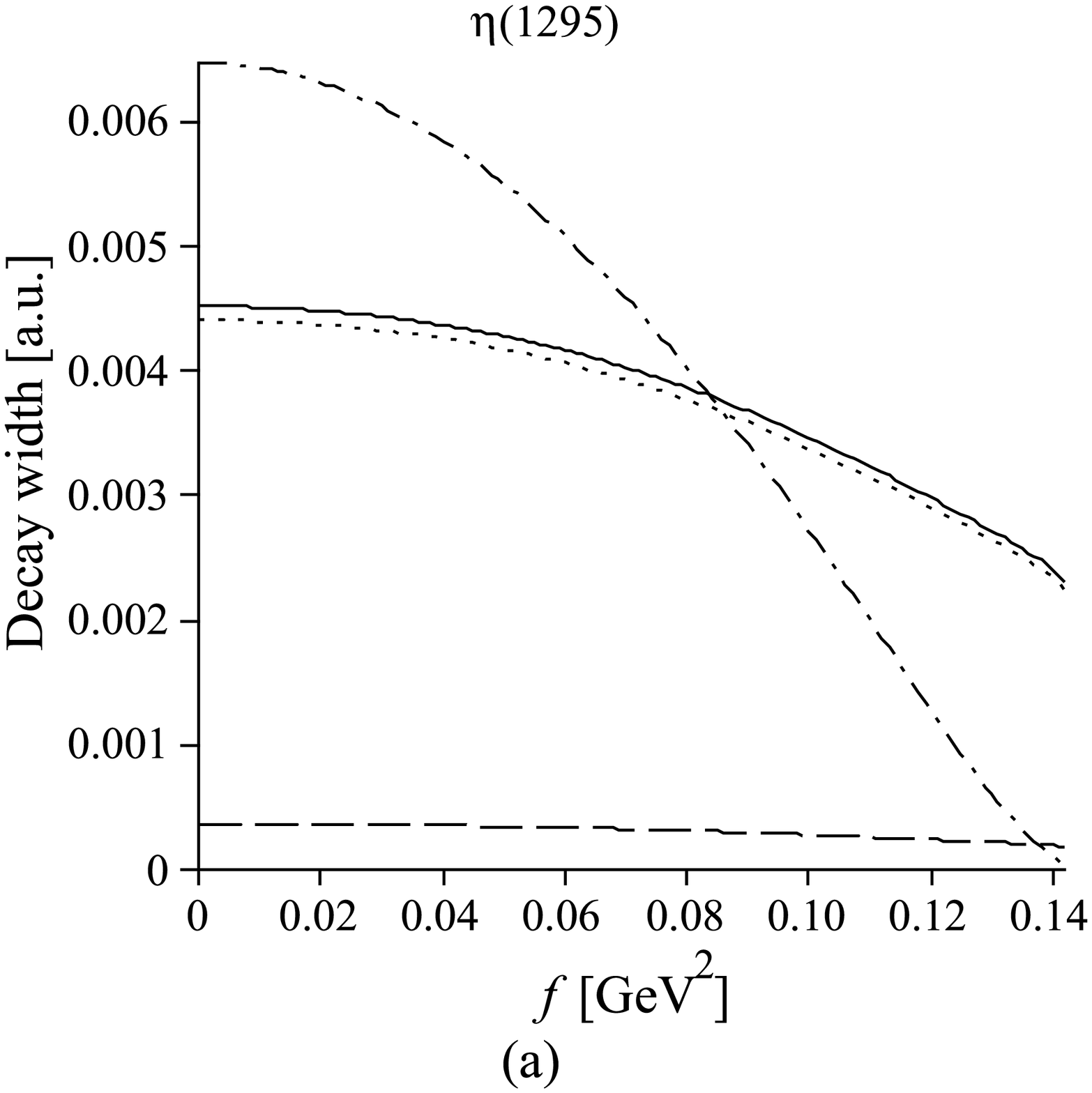} 
\includegraphics[width=7.5cm,angle=0]{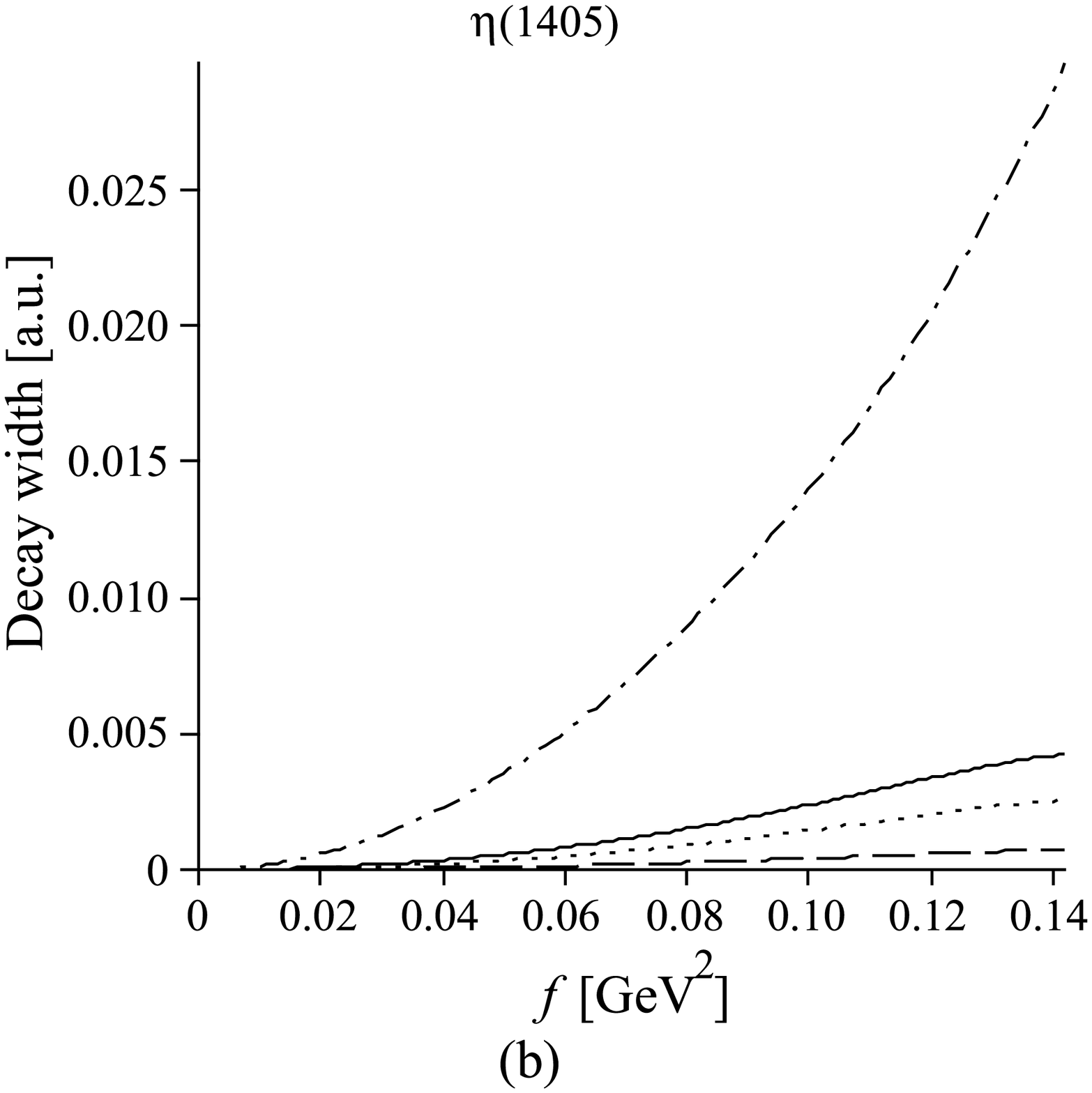} 
\includegraphics[width=7.5cm,angle=0]{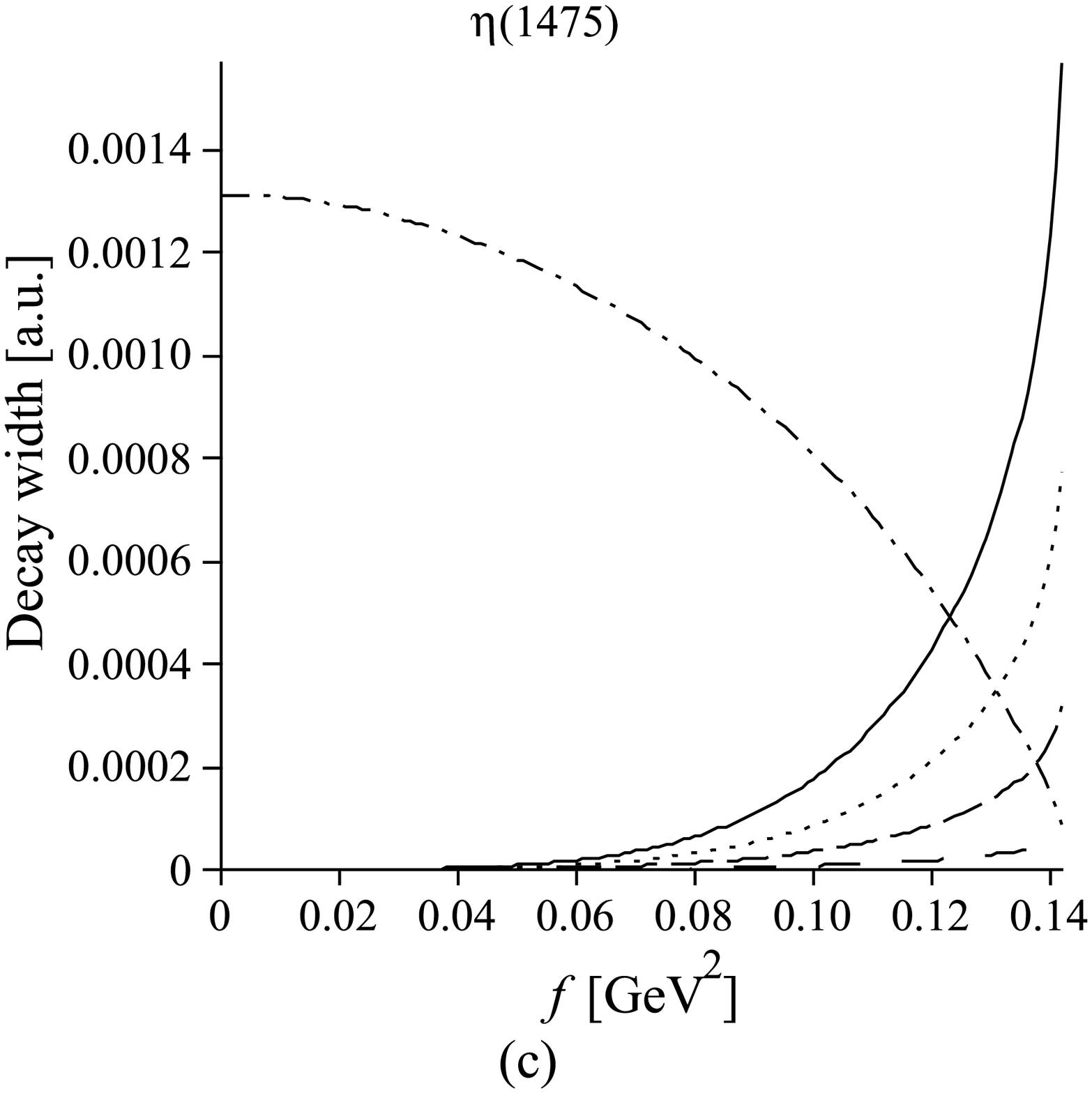} 
\end{center}
\caption{Partial decay widths of $\eta_i$ to scalar and pseudoscalar 
mesons in arbitrary normalization. Flavor symmetry breaking and 
direct glueball decay are suppressed. 
The solid line denotes decays to $a_0 \pi$, the dashed line 
$\sigma \eta^\prime$, dots denote $\sigma \eta$, the dashed-dotted 
line $\kappa K$ and the space-dashed line (only visible for 
$\eta(1475)$) denotes the decay to $f_0 \eta$. Please note that 
the dominant decay width $\Gamma(\eta(1475)\rightarrow \kappa K)$ 
has been rescaled by a factor of $0.05$ to facilitate comparison 
for convenience.}
\label{fig:scalthreesimple} 
\end{figure}

\begin{figure}
\begin{center} 
\includegraphics[width=7.5cm,angle=0]{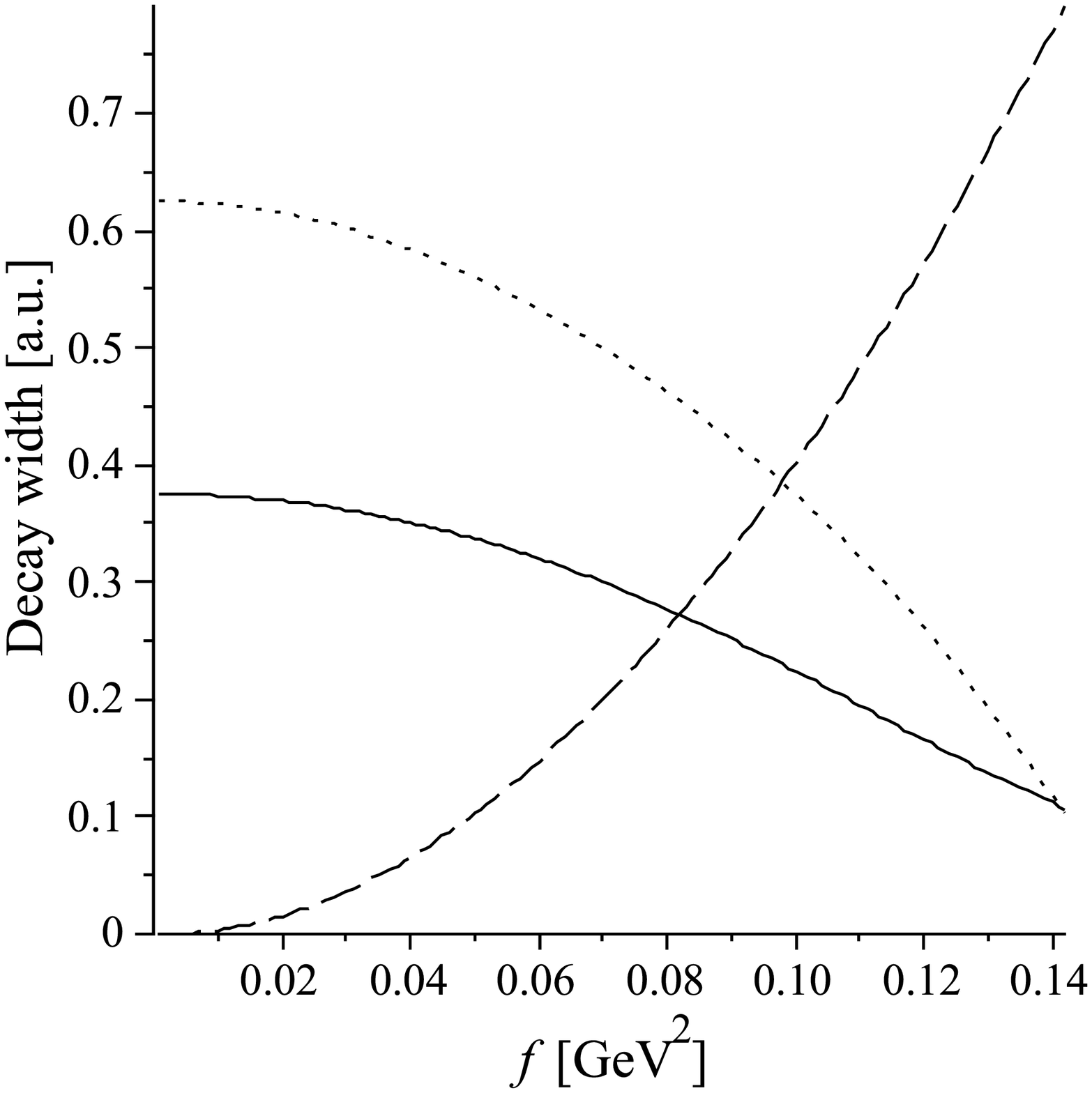} 
\end{center}
\caption{Sum of partial decay widths of $\eta_i$ to scalar and pseudoscalar 
mesons in arbitrary normalization. Flavor symmetry breaking and 
direct glueball decay are suppressed. The solid line denotes $\eta(1295)$, 
the dashed line $\eta(1405)$ and the dotted line $\eta(1475)$.}  
\label{fig:scalthreesimpletotal} 
\end{figure}

The strength of the direct glueball decay is not known a priori, and 
it is interesting to study the effect of vanishing strength on the 
decay pattern. The resonance $\eta(1405)$ can now only decay via mixing, 
its decay is strongly suppressed for small mixing strengths. The partial 
decay widths, in arbitrary normalization are shown in
 Fig.~\ref{fig:scalthreesimple}.
The total decay width to scalar and pseudoscalar mesons 
(normalized to the sum of all decay widths) 
is shown in Fig.~\ref{fig:scalthreesimpletotal}.

Three-body decays fed by intermediate resonances dominate 
in all three channels and contribute the main part to the total 
decay width of the resonances. We note that the full width of the
$\eta(1405)$ of 51.1 MeV implies that either the glueball decays directly
to scalar and pseudoscalar mesons, or the mixing with the quarkonia is 
strong, larger than $\approx 0.08$ GeV$^2$. On the other hand, if we assume 
that the mass of the bare $\eta_{n \bar n}$ is significantly smaller than the
mass of the bare $\eta_{s \bar s}$ the mixing strength has to be small 
(see Fig. \ref{baremassesforphys}). Furthermore, no  
mixing mechanism that would be strong enough is known for this sector.
While in the scalar sector the absence 
of a direct glueball decay is a realistic option based on theoretical 
arguments and might be phenomenologically successful in some cases, 
the inclusion of the direct decay is therefore needed for the scenario under 
consideration.

\subsubsection{Inclusion of direct glueball decay}

\begin{figure}
\begin{center} 
\includegraphics[width=7.2cm,angle=0]{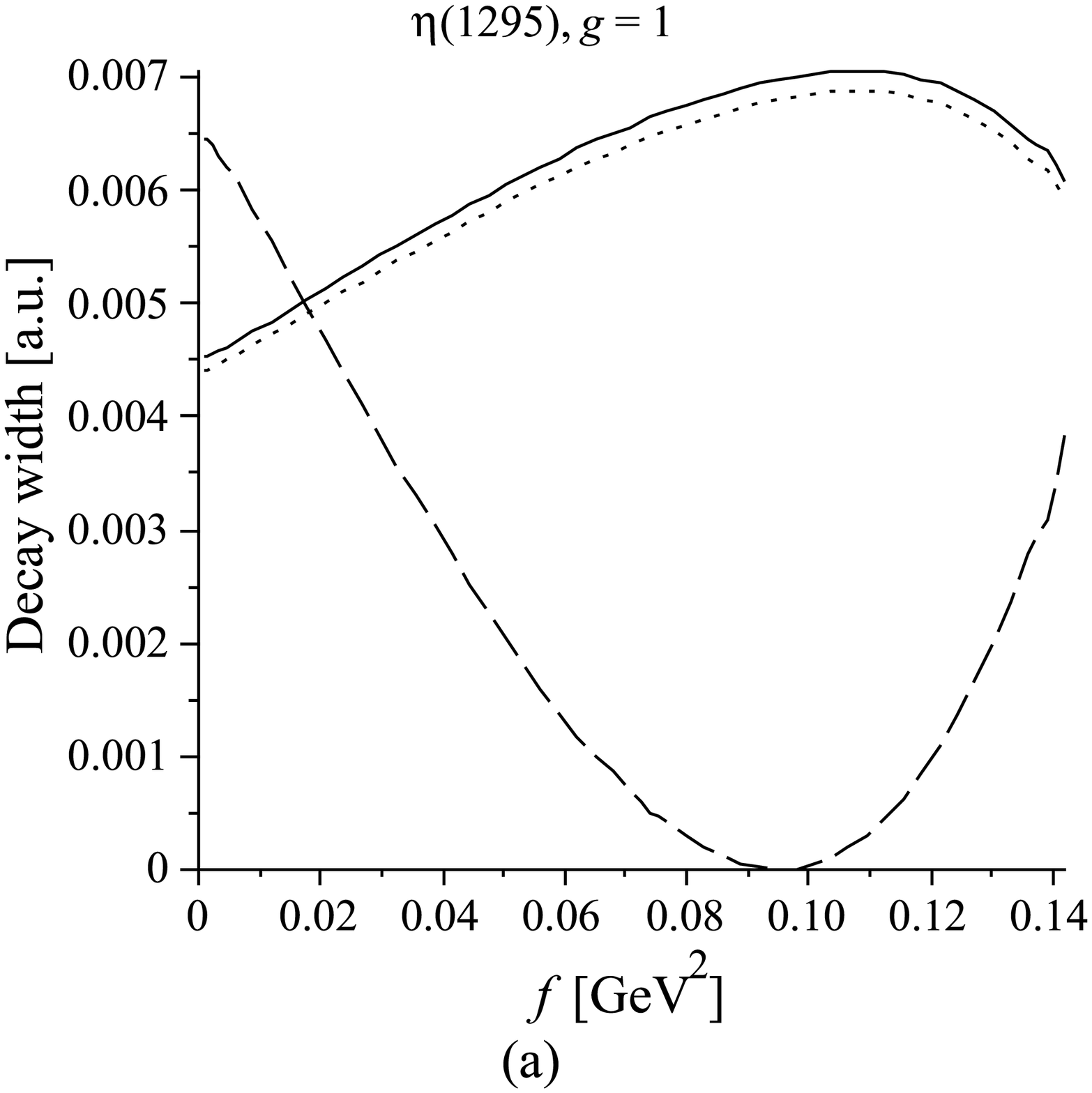}
\includegraphics[width=7.2cm,angle=0]{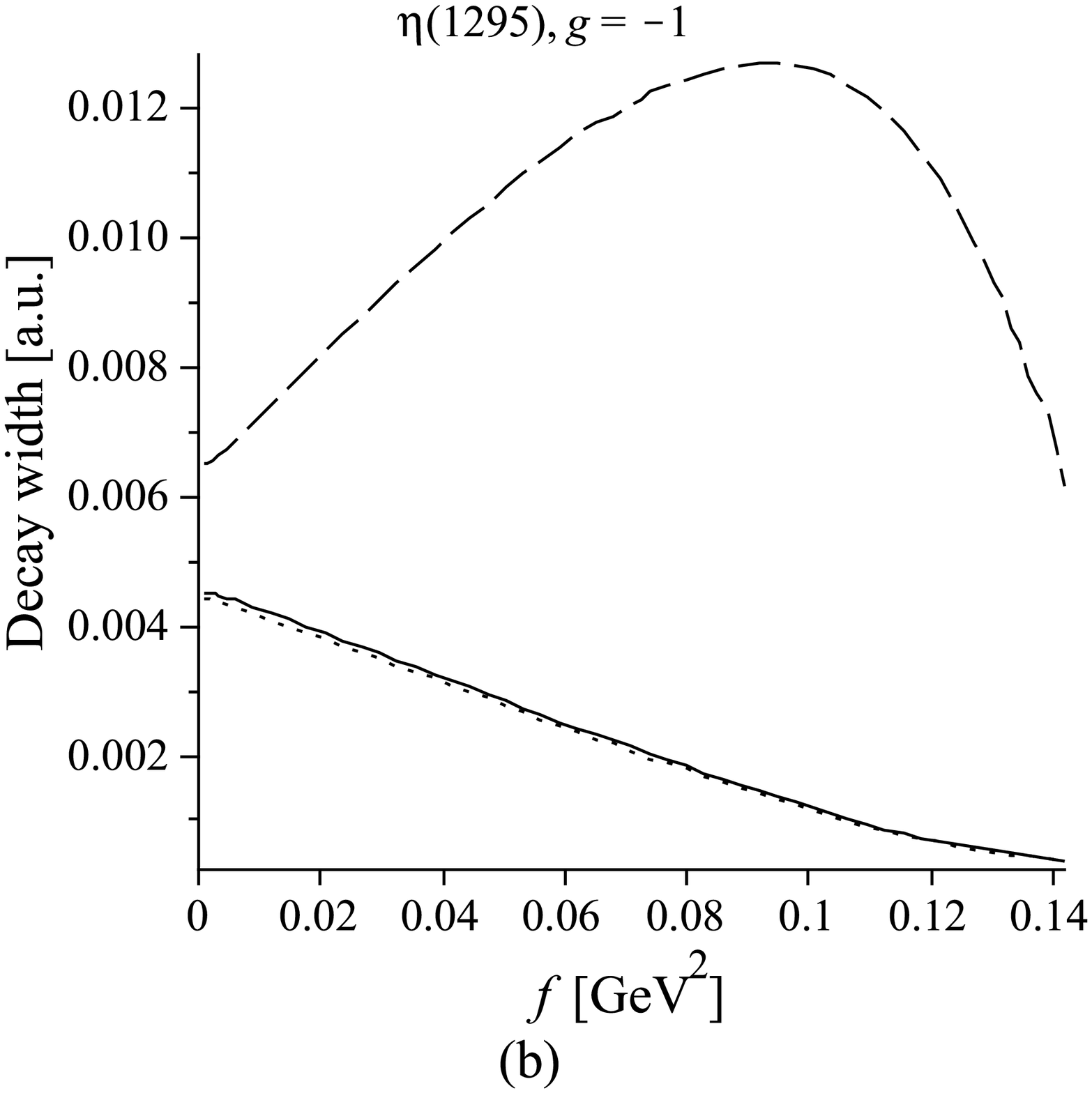}

\includegraphics[width=7.2cm,angle=0]{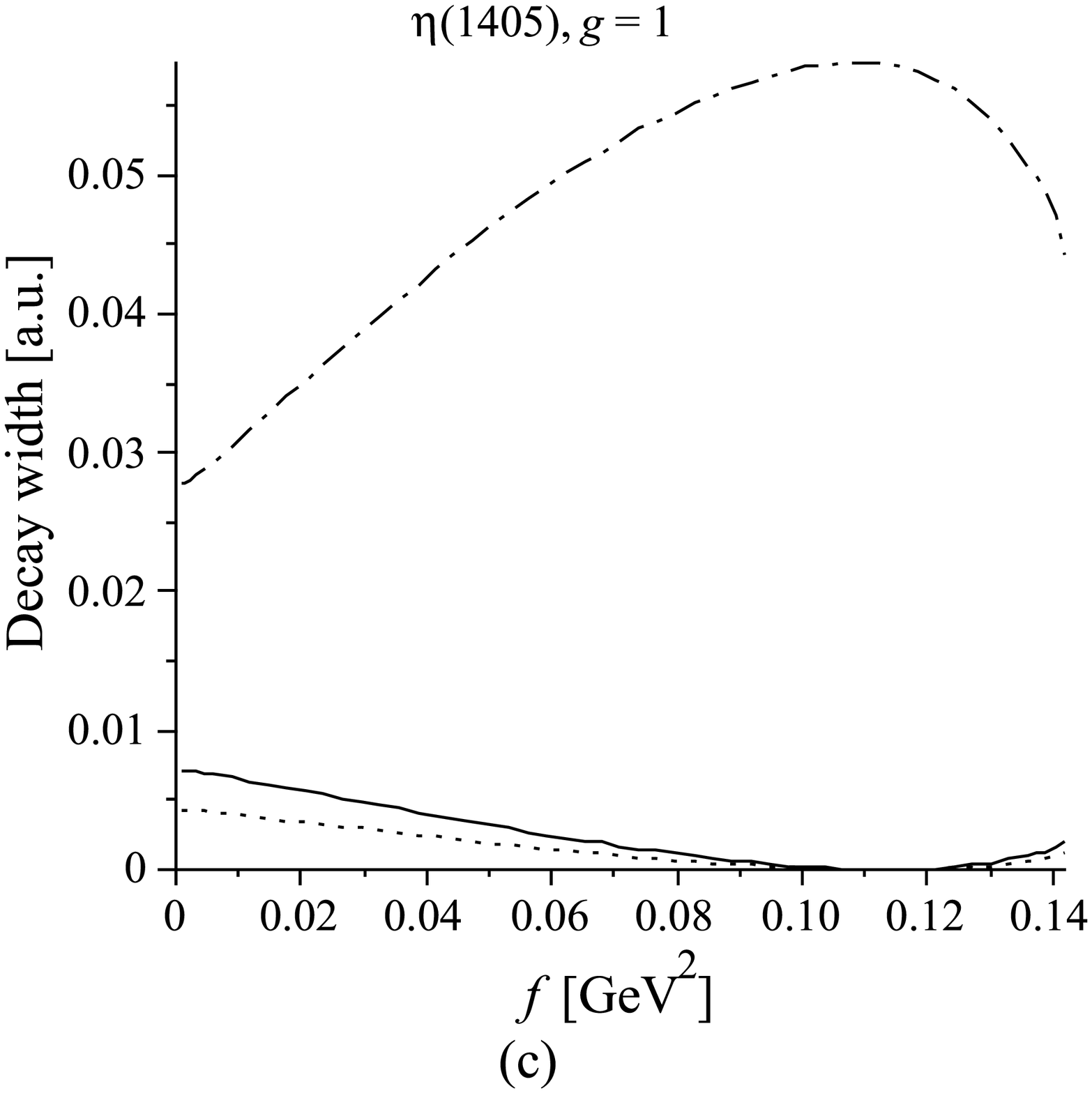}
\includegraphics[width=7.2cm,angle=0]{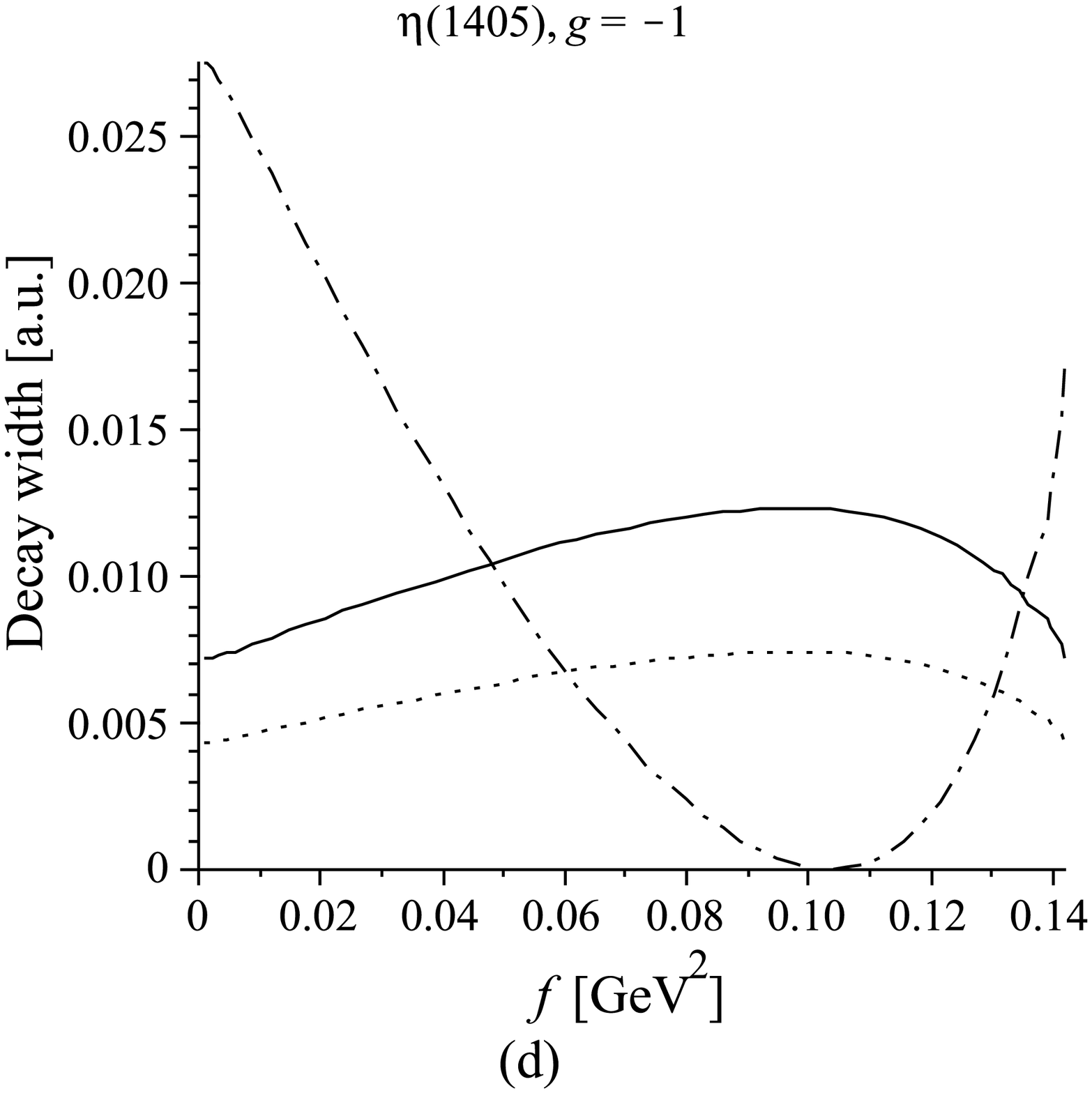}

\includegraphics[width=7.2cm,angle=0]{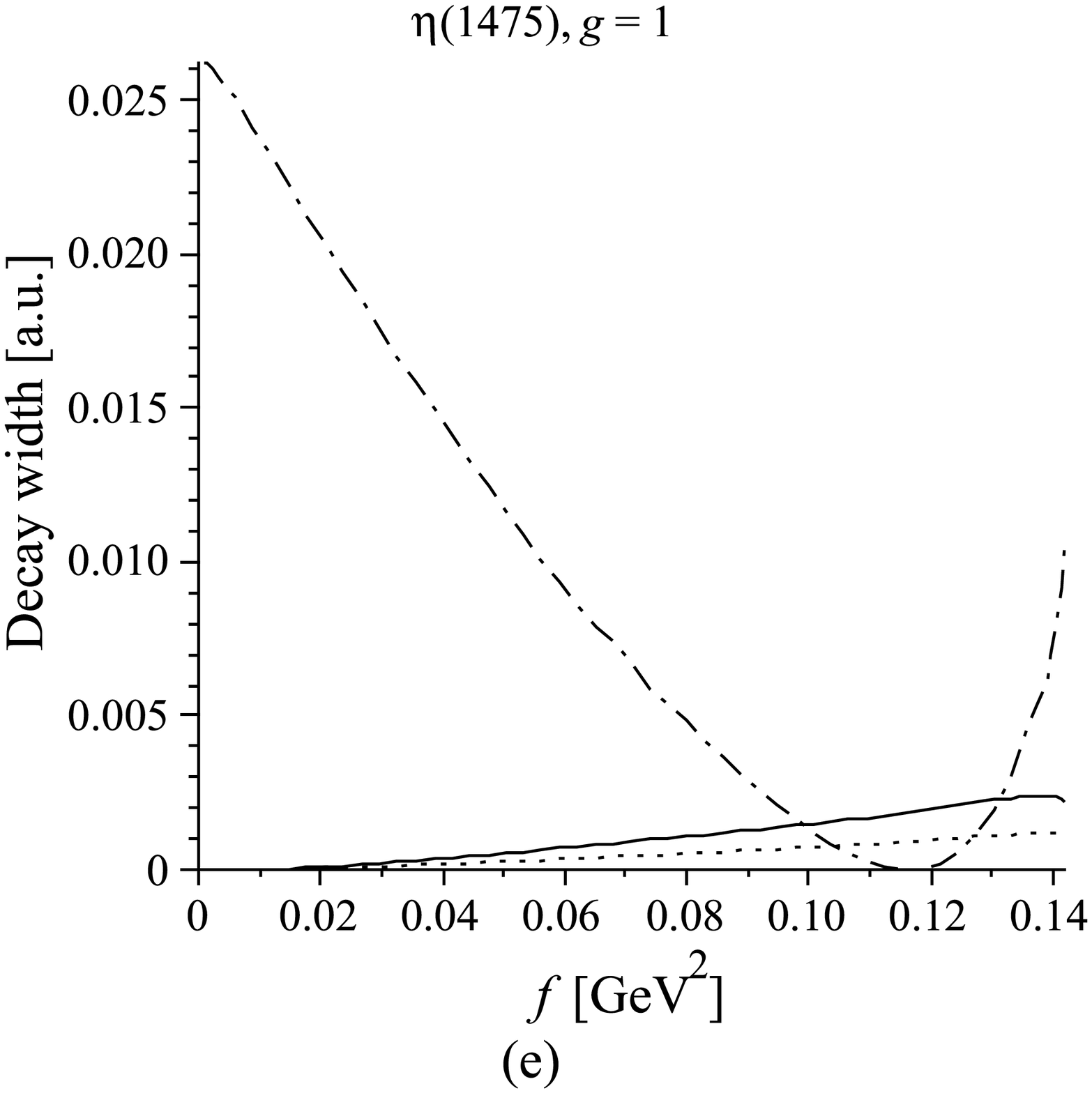}
\includegraphics[width=7.2cm,angle=0]{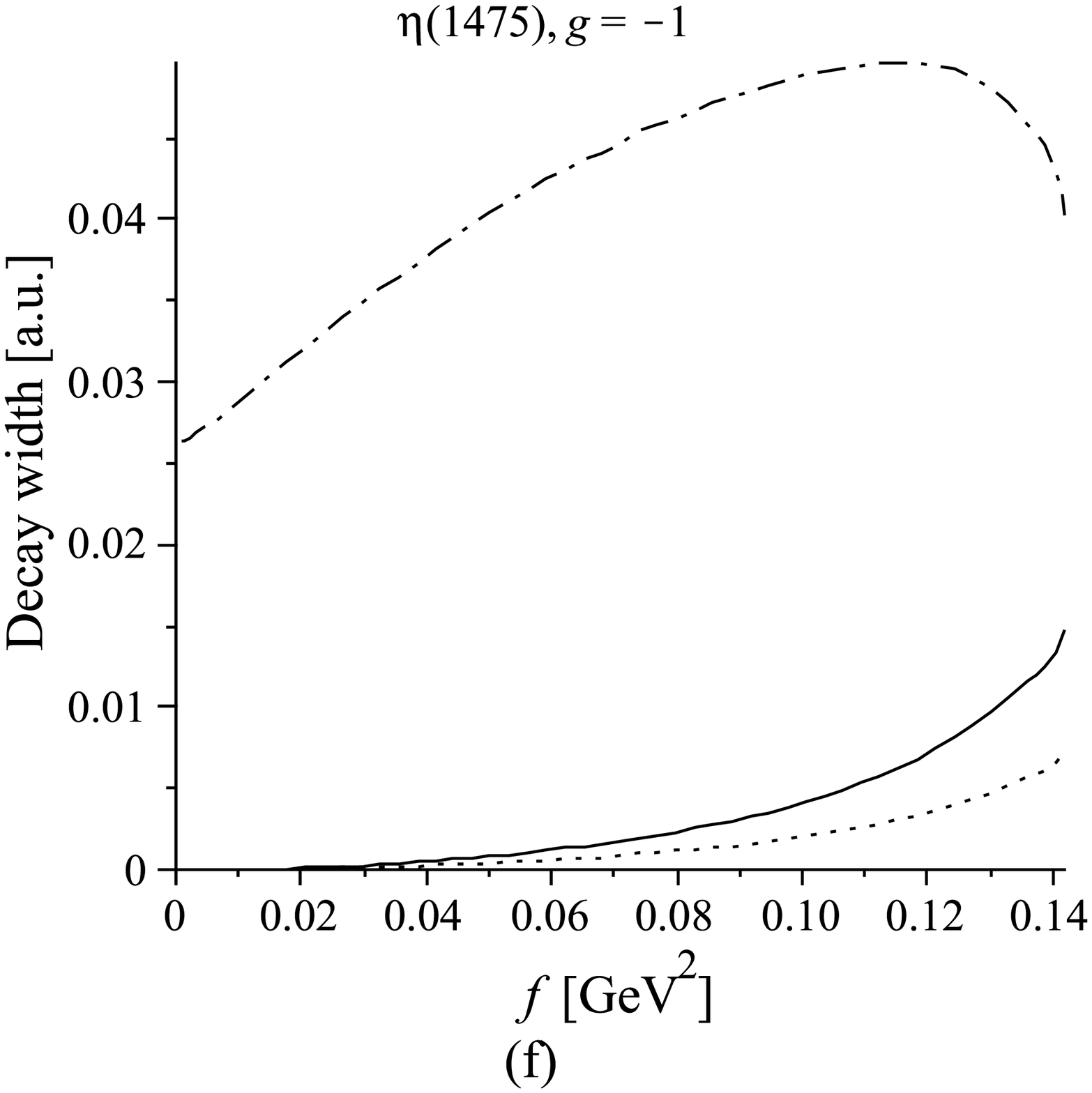}
\end{center}
\caption{Decay widths of $\eta_i$ to scalar and pseudoscalar 
mesons in arbitrary normalization including direct glueball decay. 
The solid line denotes decay to $a_0 \pi$, the dotted line denotes 
$\sigma \eta$ and the dash-dotted one $\kappa K$. 
Note the difference in the behavior of the curves for positive $g=1$ 
and negative $g=-1$ glueball decay strength.} 
\label{fig:directcompare} 
\end{figure}

The inclusion of direct glueball decay changes the picture in the case of 
the $\eta(1405)$ at low values for the mixing strength $f$ dramatically. 
The relative sign 
between the glueball and the quarkonium decay constants plays an important 
role. Negative as well as positive interference effects appear in the 
decay pattern of all three states. The decay rates for the most important 
scalar and pseudoscalar decays are shown in Fig.~\ref{fig:directcompare} 
for various choices of the direct glueball decay strength. Interference 
effects are especially dramatic in the case of the $\eta(1405)$ and 
$\eta(1295)$: While mixing leads to an increase of the decay widths of 
$\eta(1295)$ to $\sigma \eta$ and $a_0\pi$ for $g=1$, they are lowered 
for $g=-1$.  The decay mode $\eta (1475) \to \kappa K$ is totally dominant
for the case of suppressed direct glueball decay (please 
compare to Fig.~4), for intermediate values of the 
mixing angles other decay modes begin to be important.

Current data on the $a_0 \pi$ and $\sigma \eta$ (or $(\pi \pi)_{S-wave} \eta$)
decay channels are available for the $\eta (1295)$ and $\eta (1405)$.
The Crystal Barrel Collaboration~\cite{Amsler:1995wz} reports a value 
for the ratio 
\eq 
\mbox{Br}( \eta (1405) \to \eta \sigma)/ \mbox{Br} ( \eta (1405) 
\to a_0 \pi ,~a_0 \to \eta \pi)=
0.78\pm 0.12\pm 0.10 \,, 
\en 
consistent with the inverse ratio
\eq 
\mbox{Br}( \eta (1405) \to a_0 \pi)/ 
\mbox{Br}(\eta (1405) \to \eta \sigma) =
0.91\pm 0.12 
\en 
of Ref.~\cite{Anisovich:2001jb}. 
These values should be compared
to the BES result~\cite{Bai:1999tg} of 
\eq 
\mbox{Br}( \eta (1405) \to a_0 \pi)/ \mbox{Br}( \eta (1405)
\to \eta \sigma) = 0.70\pm 0.12\pm 0.20 \,, 
\en 
The E852 Collaboration published a value of 
\eq 
\mbox{Br} ( \eta (1405) \to a_0 \pi)/ \mbox{Br}( \eta (1405)
\to \eta \sigma) = 0.15\pm 0.04 
\en 
in~\cite{Manak:2000px}, statistical error indicated only,
in conflict with above mentioned values.
Whereas first analyses indicate that the $a_0 \pi$ and $\sigma \eta$ 
decay modes 
are roughly of equal strength, the E852 result implies a dominant $\eta \sigma$
mode. Last scenario cannot be reproduced in the present model, even when
the value of direct glueball decay strength $g$ is increased dramatically. 
From our predictions we always find that $a_0 \pi$ dominates with varying 
strength over $\sigma \eta$.  This effect is traced to the fact that we place 
$\sigma$ and $\kappa$ in the chiral formalism in the same flavor octet.

For the $\eta (1295)$ the GAMS Collaboration~\cite{Alde:1996kx} reports 
the ratio 
\eq 
\mbox{Br}(\eta (1295) \to \eta \sigma)/ \mbox{Br}(\eta (1295) 
\to a_0 \pi, a_0 \to \eta \pi) = 0.54 \pm 0.22 \,. 
\en  
The E852 collaboration extract from their analysis the 
result~\cite{Manak:2000px} 
\eq 
\mbox{Br}(\eta (1295) \to a_0 \pi)/\mbox{Br}(\eta (1295) 
\to \eta \sigma) = 0.48\pm 0.22 \,. 
\en 
Similar to the behaviour of the $\eta (1405)$ for the $\eta (1295)$ 
we deduce that the $a_0 \pi$ mode slightly dominates over $\sigma \eta$, 
nearly independent of values for the mixing and glueball decay strength.

\subsection{Discussion}

We can safely assume that the bare $\eta_{nn}$ mass is considerably 
lighter than the mass of $\eta_{ss}$. To be able to explain the mass values
of the $\eta$-states we furthermore
have to restrict to a small glueball-quarkonia mixing strength.
For these small values of the mixing strength and, 
in addition, a vanishing direct glueball decay, we expect the width of 
the $\eta(1405)$ to be much smaller than the values for $\eta(1295)$ and 
$\eta(1475)$. This is however not the case, since 
$\Gamma(\eta(1405)) \approx 51$ MeV which is comparable to the total width 
of $\eta(1295)$. While in the scalar sector the direct glueball decay 
process may be suppressed, it has to play a dominant role in the pseudoscalar 
channel in order to explain the width of the $\eta(1405)$.

A Dalitz plot analysis of the $\pi K \overline K$ and $\pi \pi \eta$ mode 
in $\eta(1405)$ would be interesting for various reasons: For a very 
small mixing strength the decays to $K \overline K^\ast$ are suppressed 
since these channels are not fed by the direct glueball decay. Therefore, 
the decay to $K K \pi$ cannot proceed by an intermediate vector resonance, 
an enhancement would be a signal for considerable $q \bar q$ admixture 
if it cannot be attributed to $\kappa K$. Similarly, we may extract 
information on the glueball- or $n \bar n$-content of the $\eta(1475)$ 
by the search for the decay modes $a_0 \pi, \sigma \eta$ in three-body 
decays. The measurement of $K \overline K^\ast$ in $\eta(1405)$ would 
give a good estimate of the possible $q \bar q$-components and the 
mixing strength.

The situation is more difficult when trying to estimate the glueball or 
$s \bar s$ admixture in the $\eta(1295)$ since the decay modes are all 
open and are only slightly modified when varying the mixing strength. 
The actually known decay pattern and its consequence for the 
interpretation of the pseudoscalar mesons is the following:
\begin{itemize}
\item The dominant decay modes for $\eta(1475)$ are $K \overline K^\ast$ 
and $\kappa K$, which are revealed by the $K\overline K \pi$ final state 
in the experiment. A Dalitz plot analysis of the $K K \pi$ mode is 
important to extract the scalar ($\kappa$) and vector meson ($K^\ast$) 
resonance contributions. Non-strange decay modes are strongly suppressed 
and their observation would point to a glueball-component or 
$n \bar n$-configuration. The actual observation of $a_0 \pi$, although 
relatively weak, points to some glueball admixture. The strong dominance 
of $K K \pi$ is compatible with our model, in which $\kappa K$ dominates 
the scalar-pseudoscalar channel and $K \overline{K^\ast}$ the 
vector-pseudoscalar channel.
\item We expect the decays to scalar and pseusoscalar mesons for the 
$\eta(1295)$ to be dominant, the $K \overline K^\ast$ being suppressed 
kinematically. The decays to $K \overline K \pi$ and $\pi \pi \eta$ do 
not depend strongly on the mixing strength.
\item For reasonable mixing strengths, $K \overline K^\ast$ is suppressed 
for $\eta(1405)$ and we expect the $K \overline K \pi$ to mainly arise 
from scalar resonances. The decays to $\rho \gamma$ and $\phi \gamma$, 
which have been observed according to PDG~\cite{Amsler:2008zz}, 
are compatible with a glueball interpretation as well as the decay to 
$a_0 \pi$. On the other hand, $K \overline{ K^\ast}$ and $\gamma \gamma$ 
are weak in comparison to the decays into scalar and pseudoscalars. 
Our model shows that their bare observation points to a non-vanishing 
$s \bar s$- or $n \bar n$-component. 
\end{itemize}

\section{Conclusions and Outlook} \label{last}

Qualitatively, the scenario we have considered shows reasonable agreement 
with the decay processes observed so far. Further experimental input 
would be appreciated, especially for the $\eta(1405)$ and $\eta(1475)$. 
The glueball contribution to $\eta(1295)$ should be approximately as 
large as the contribution to $\eta(1475)$, but it is reflected more 
strongly in the decay pattern of the latter. Similarly, an 
$s \bar s$-admixture to $\eta(1405)$ would be as large as the 
$n \bar n$-admixture, but would lead to greater changes in the 
decay modes. The dominant $n \bar n$-structure of $\eta(1295)$ 
is rather well established, especially due to the mass degeneracy 
with the $\pi(1300)$. Quantitative determination of glueball and 
$s \bar s$-admixtures are difficult since no new channels open up 
when mixing is included; only subtle changes in the decay pattern 
would be observable. On the other hand, the observation of 
non-strange decay modes in $\eta(1475)$ or further strange decays 
such as $\kappa K$ in $\eta(1405)$-decays would point to considerable 
mixing in this region and help to quantify it. 

On the experimental side, exciting results can be expected in the next 
years: planned experiments at BES-III, COMPASS and at the upgrade facility 
FAIR at GSI might give essential contributions to map out the decay modes 
of the $\eta$-states~\cite{Crede:2008vw}.

If the glueball interpretation was confirmed by experiment, the first 
question which naturally arises would be why the theoretical predictions 
for the $J^{\rm PC} = 0^{-+}$ glueball are dominantly in a very different 
mass region. We have seen that mixing cannot change the physical mass of 
the glueball dramatically and hence cannot offer an explanation for 
the strong deviation from the mainstream theoretical predictions.

The question whether the $\eta(1405)$ may be the lowest pseudoscalar 
glueball cannot be answered conclusively at this stage. We have shown 
however that the scenario under consideration is in qualitative agreement 
with the available experimental data. We have also pointed out how the 
problem of the pseudoscalar glueball may be solved in the future at 
a more quantitative level. 

\begin{acknowledgments}

We would like to thank Alberto Masoni and Claude Amsler for helpful
discussions on experimental issues.
This work was supported by the DFG under Contract No. FA67/31-1, 
No. FA67/31-2, and No. GRK683. This research is also part of
the European Community-Research Infrastructure Integrating Activity
"Study of Strongly Interacting Matter" (HadronPhysics2, Grant Agreement 
n. 227431) and of the President grant of Russia
"Scientific Schools"  No. 871.2008.2. 

\end{acknowledgments}

\appendix
\label{appen}
\section{Matrices   $\cal S$, $\cal{V}_{\mu\nu}$ and $\cal{P}^\ast$}

\eq 
\cal S &=&  \left(\begin{array}{ccc} \displaystyle\frac{a_0}{\sqrt 2} 
+ \displaystyle\frac{\sigma}{\sqrt 2} & a_0^+ & \kappa^+    \\  
a_0^- & -\displaystyle\frac{a_0}{\sqrt 2} 
+ \displaystyle\frac{\sigma}{\sqrt 2} &\kappa^0 \\
\kappa^-& \overline{\kappa^0} & f_0\end{array}    \right) \,, \\[3mm]   
\cal{V}_{\mu\nu} &=& 
\left(\begin{array}{ccc}
 \displaystyle\frac{\rho^0}{\sqrt{2}} 
+ \displaystyle\frac{\omega}{\sqrt{2}}  
& \rho^+ & K^{\ast \, +} \\ 
 \rho^-& - \displaystyle\frac{\rho_0}{\sqrt{2}} 
+ \displaystyle\frac{\omega}{\sqrt{2}} & K^{\ast 0}\\
 K^{\ast \, -} & \overline{K^{\ast \, 0}} & \phi  
 \end{array}\right)_{\mu\nu} \,, \\[3mm]  
\cal{P}^\ast &=& \left(\begin{array}{ccc}
  \displaystyle\frac{\pi^0(1300)}{\sqrt{2}} + \frac{\eta(1295)}{\sqrt{2}} 
& \pi^+(1300) & K^+(1460)\\
 \pi^-(1300)& - \displaystyle\frac{\pi^0(1300)}{\sqrt{2}} 
 + \frac{\eta(1295)}{\sqrt{2}} & K^0(1460)\\
 K^-(1460)& \overline{K^0}(1460) &  \eta(1475) 
 \end{array}\right)  
\en


\begin{thebibliography}{99} 

\bibitem{Donoghue:1980hw}
  J.~F.~Donoghue, K.~Johnson and B.~A.~Li,
  Phys.\ Lett.\  B {\bf 99}, 416 (1981); 
  K.~Ishikawa, 
  Phys.\ Rev.\ Lett.\  {\bf 46}, 978 (1981); 
  M.~S.~Chanowitz,
  Phys.\ Rev.\ Lett.\  {\bf 46}, 981 (1981);  
  R.~Lacaze and H.~Navelet,
  Nucl.\ Phys.\  B {\bf 186}, 247 (1981); 
  C.~E.~Carlson, J.~J.~Coyne, P.~M.~Fishbane, F.~Gross and S.~Meshkov,
  Phys.\ Lett.\  B {\bf 98}, 110 (1981);  
  M.~A.~Ivanov and R.~K.~Muradov,
  JETP Lett.\  {\bf 42}, 367 (1985).
\bibitem{Chen:2005mg}
  Y.~Chen {\it et al.}, 
  Phys. \ Rev. \ D {\bf 73}, 014516 (2006).
\bibitem{Meyer:2004jc}
  H.~B.~Meyer and M.~J.~Teper, 
  Phys. \ Lett. \ B {\bf 605}, 344 (2005).
\bibitem{Bai:1990hk}
  Z.~Bai {\it et al.}, 
  Phys. \ Rev. \ Lett. {\bf 65}, 1309 (1990).
\bibitem{Heusch:1991sw}
  C.~A. Heusch.
  Lecture at the 5th Rencontres De Physique De La Vallee D'Aoste, La
  Thuile, Italy, Mar 3-9, 1991: Results and Perspectives in Particle Physics
  Proceedings. - Ed. Frontieres, Gif-Sur-Yvette, 1991.
\bibitem{Kochelev:2005tu}
  N.~Kochelev and D.~P.~Min,
  Phys.\ Rev.\  D {\bf 72}, 097502 (2005); 
  Phys.\ Lett.\  B {\bf 633}, 283 (2006). 
\bibitem{Thomas:2007uy}
  C.~E.~Thomas, 
  JHEP {\bf 10}, 026 (2007).
\bibitem{Ambrosino:2006gk}
  F.~Ambrosino {\it et al.},  
  Phys. \ Lett. \ B {\bf 648}, 267 (2007).
\bibitem{Escribano:2007cd}
  R.~Escribano and J.~Nadal, 
  JHEP {\bf 05}, 006 (2007). 
\bibitem{Cheng:2008ss}
  H.~Y.~Cheng, H.~N.~Li and K.~F.~Liu,
  Phys.\ Rev.\  D {\bf 79}, 014024 (2009). 
\bibitem{Masoni:2006rz}
  A.~Masoni, C.~Cicalo, and G.~L.~Usai.
  J. \ Phys.\ G {\bf 32}, R293 (2006).
\bibitem{Gabadadze:1997zc}
  G.~Gabadadze, 
  Phys. \ Rev. \ D {\bf 58}, 055003 (1998). 
\bibitem{Morningstar:1999rf}
  C.~J.~Morningstar and M.~J.~Peardon,
  Phys.\ Rev.\  D {\bf 60}, 034509 (1999).  
\bibitem{Faddeev:2003aw}
  L.~Faddeev, A.~J.~Niemi, and U.~Wiedner, 
  Phys. \ Rev. \ D {\bf 70}, 114033 (2004).
\bibitem{He:2009sb}
 S.~He, M.~Huang and Q.~S.~Yan,
arXiv:0903.5032 [hep-ph].
\bibitem{Fariborz1}
A.~H.~Fariborz, R.~Jora and J.~Schechter, Int. J. Mod. Phys. A{\bf 20}, 6178 (2005)
\bibitem{Fariborz2}
  A.~H.~Fariborz, R.~Jora and J.~Schechter,
     arXiv:0902.2825 [hep-ph]
\bibitem{Close:1987er}
F.~E.~Close, 
  Rept. \ Prog. \ Phys. {\bf 51}, 833 (1988).
\bibitem{Bali:1993fb}
  G.~S.~Bali, K.~Schilling, A.~Hulsebos, 
  A.~C.~Irving, C.~Michael and P.~W.~Stephenson [UKQCD Collaboration],
  Phys.\ Lett.\  B {\bf 309}, 378 (1993). 
\bibitem{Amsler:1995td}
  C.~Amsler and F.~E~ Close, 
  Phys. \ Rev. \ D {\bf 53}, 295 (1996).  
\bibitem{Narison:1996fm}
  S.~Narison,
  Nucl.\ Phys.\  B {\bf 509}, 312 (1998). 
\bibitem{Feldmann:1999uf}
  T.~Feldmann,
  Int.\ J.\ Mod.\ Phys.\  A {\bf 15}, 159 (2000). 
\bibitem{Ebert:2000nx}
  D.~Ebert, M.~Nagy, M.~K.~Volkov, and V.~L.~Yudichev,
  Eur.\ Phys.\ J.\  A {\bf 8}, 567 (2000). 
\bibitem{Volkov:2001ct}
  M.~K.~Volkov and V.~L.~Yudichev,
  Eur.\ Phys.\ J.\  A {\bf 10}, 223 (2001). 
\bibitem{Burdanov:2000rw}
  J.~V.~Burdanov and G.~V.~Efimov,
  Phys.\ Rev.\  D {\bf 64}, 014001 (2001), 
\bibitem{Giacosa:2004ug}
  F.~Giacosa, T.~Gutsche, and A.~Faessler,
  Phys.\ Rev.\  C {\bf 71}, 025202 (2005). 
\bibitem{Giacosa:2005qr}
  F.~Giacosa, T.~Gutsche, V.~E.~Lyubovitskij, and A.~Faessler,
  Phys.\ Lett.\  B {\bf 622}, 277 (2005); 
  Phys.\ Rev.\  D {\bf 72}, 094006 (2005). 
\bibitem{Giacosa:2005bw}
  F.~Giacosa, T.~Gutsche, V.~E.~Lyubovitskij, and A.~Faessler, 
  Phys.\ Rev.\  D {\bf 72}, 114021 (2005). 
\bibitem{Forkel:2003mk}
H.~Forkel, 
  Phys.\ Rev.\  D {\bf 71}, 054008 (2005). 
\bibitem{Cheng:2006hu}
  H.~Y.~Cheng, C.~K.~Chua and K.~F.~Liu, 
  Phys.\ Rev.\  D {\bf 74}, 094005 (2006). 
\bibitem{Giacosa:2006tf}
  F.~Giacosa, 
  Phys. \ Rev. \ D {\bf 75}, 054007 (2007).
\bibitem{Gerasimov:2007sb}
  S.~B.~Gerasimov, M.~Majewski, and V.~A.~Meshcheryakov, 
  arXiv:0708.3762 [hep-ph].
\bibitem{Klempt:2007cp}
  E.~Klempt and A.~Zaitsev.
  Phys. \ Rept.\ {\bf 454}, 1 (2007). 
\bibitem{Mathieu:2008me}
  V.~Mathieu, N.~Kochelev and V.~Vento,
  Int.\ J.\ Mod.\ Phys.\  E {\bf 18}, 1 (2009). 
\bibitem{Mathieu:2008up}
  V.~Mathieu, F.~Buisseret, C.~Semay and B.~Silvestre-Brac,
  arXiv:0811.2710 [hep-ph].
\bibitem{Weinberg:1978kz}
  S.~Weinberg,
  Physica \ A {\bf 96}, 327 (1979). 
\bibitem{Gasser:1983yg}
  J.~Gasser and H.~Leutwyler,
  Annals Phys.\  {\bf 158}, 142 (1984); 
  Nucl.\ Phys.\ B {\bf 250}, 465 (1985).
\bibitem{Ecker:1988te} G.~Ecker, J.~Gasser, A.~Pich, and E.~de Rafael, 
  Nucl.\ Phys.\ B {\bf 321}, 311 (1989); 
  G.~Ecker, J.~Gasser, H.~Leutwyler, A.~Pich, and E.~de Rafael, 
  Phys.\ Lett.\ B {\bf 223}, 425 (1989). 
\bibitem{excited_mesons}
T.~Gutsche, V.~E. Lyubovitskij, and M.~C.~Tichy, 
Phys.\ Rev. \ D {\bf 79}, 014036 (2009).
\bibitem{Li:2007ky}
  G.~Li, Q.~Zhao and C.~H.~Chang,
        J.\ Phys.\ G {\bf 35}, 055002 (2008).
\bibitem{Crede:2008vw}
  V.~Crede and C.~A.~Meyer,
  %``The Experimental Status of Glueballs,''
  Prog.\ Part.\ Nucl.\ Phys.\  {\bf 63}, 74 (2009). 
\bibitem{Bettoni:2005ut}
  D.~Bettoni, 
  J. \ Phys. \ Conf. \ Ser. {\bf 9}, 309 (2005). 
\bibitem{Lee:1999kv}
    W.~J.~Lee and D.~Weingarten,
        Phys.\ Rev.\  D {\bf 61}, 014015 (2000).
\bibitem{Barnes:1996ff}
  T.~Barnes, F.~E. Close, P.~R. Page, and E.~S. Swanson, 
  Phys. \ Rev. \ D {\bf 55}, 4157 (1997). 
\bibitem{Barnes:2002mu}
  T.~Barnes, N.~Black, and P.~R.~Page, 
  Phys. \ Rev. \ D {\bf 68}, 054014 (2003). 
 \bibitem{Amsler:2008zz}
  C.~Amsler {\it et al.}  [Particle Data Group], 
  Phys.\ Lett.\  B {\bf 667}, 1 (2008). 
  \bibitem{Adams:2001sk}
    G.~S.~Adams {\it et al.}  [E852 Collaboration],
  Phys.\ Lett.\  B {\bf 516}, 264 (2001).
  \bibitem{Amsler:1995wz}
    C.~Amsler {\it et al.}  [Crystal Barrel Collaboration],
        Phys.\ Lett.\  B {\bf 358}, 389 (1995).
  \bibitem{Anisovich:2001jb}
    A.~V.~Anisovich {\it et al.},
      Nucl.\ Phys.\  A {\bf 690}, 567 (2001).
 \bibitem{Bai:1999tg}
        J.~Z.~Bai {\it et al.}  [BES Collaboration],
   Phys.\ Lett.\  B {\bf 446}, 356 (1999).
   \bibitem{Manak:2000px}
     J.~J.~Manak {\it et al.}  [E852 Collaboration],
	   Phys.\ Rev.\  D {\bf 62}, 012003 (2000).
\bibitem{Alde:1996kx}
  D.~Alde {\it et al.}  [GAMS Collaboration],
  Phys.\ Atom.\ Nucl.\  {\bf 60}, 386 (1997)
    [Yad.\ Fiz.\  {\bf 60}, 458 (1997)].

\end{thebibliography}
\end{document}